\documentclass{emulateapj}

\usepackage{graphicx}
\usepackage{natbib}
\usepackage{color}
\usepackage{float}
\usepackage{txfonts}

\newcommand{\m}{$^{\rm m}\!\!.$}

\begin{document}

\title{First study of 54 new eccentric eclipsing binaries in our Galaxy}

\author{Zasche, P.~\altaffilmark{1},
     Wolf, M.~\altaffilmark{1},
     Uhla\v{r}, R.~\altaffilmark{2},
     Caga\v{s}, P.~\altaffilmark{3},
     Jury\v{s}ek, J.~\altaffilmark{4},
     Ma\v{s}ek, M.~\altaffilmark{4,5},
     Ho\v{n}kov\'a, K.~\altaffilmark{5},
     Ku\v{c}\'akov\'a, H.~\altaffilmark{1},
     Lehk\'y, M.~\altaffilmark{6,7},
     Kotkov\'a, L.~\altaffilmark{8},
     White, G.J.~\altaffilmark{9,10},
     Bewsher, D.~\altaffilmark{11},
     Tyl\v{s}ar, M.~\altaffilmark{12},
     Jel\'{\i}nek, M.~\altaffilmark{8,13},
     Paschke, A.~\altaffilmark{14} }

\email{zasche@sirrah.troja.mff.cuni.cz}

 \affil{
  \altaffilmark{1} Astronomical Institute, Charles University, Faculty of Mathematics and Physics, CZ-180~00, Praha 8, V~Hole\v{s}ovi\v{c}k\'ach 2, Czech Republic, \\
  \altaffilmark{2} Private Observatory, Poho\v{r}\'{\i} 71, CZ-254~01, J\'{\i}lov\'e u Prahy, Czech Republic, \\
  \altaffilmark{3} BSObservatory, Modr\'a 587, CZ-760~01, Zl\'{\i}n, Czech Republic, \\
  \altaffilmark{4} Institute of Physics, Czech Academy of Sciences, Na Slovance 1999/2, CZ-182~21, Praha 8, Czech Republic, \\
  \altaffilmark{5} Variable Star and Exoplanet Section of the Czech Astronomical Society, Vset\'{\i}nsk\'a 941/78, CZ-757 01 Vala\v{s}sk\'e Mezi\v{r}\'{\i}\v{c}\'{\i}, Czech Republic, \\
  \altaffilmark{6} \'Upice Observatory, U Lipek 160, CZ-542 32, \'Upice, Czech Republic, \\
  \altaffilmark{7} Astronomical Society Hradec Kr\'alov\'e, Z\'ame\v{c}ek 456/30, CZ-500 08 Hradec Kr\'alov\'e, Czech Republic, \\
  \altaffilmark{8} Astronomical Institute, The Czech Academy of Sciences, Fri\v{c}ova 298, CZ-251 65, Ond\v{r}ejov, Czech Republic, \\
  \altaffilmark{9} School of Physical Sciences, The Open University, Walton Hall, Milton Keynes MK6 7AA, UK, \\
  \altaffilmark{10} RAL Space, The Rutherford Appleton Laboratory, Chilton, Didcot, Oxfordshire OX11 0NL, UK, \\
  \altaffilmark{11} Jeremiah Horrocks Institute, University of Central Lancashire, Preston PR1 2HE, UK, \\
  \altaffilmark{12} Hv\v{e}zd\'arna, Kol\'a\v{r}ovy sady 3348, CZ-796 01 Prost\v{e}jov, \\
  \altaffilmark{13} Instituto de Astrof\'{\i}sica de Andaluc\'{\i}a, P.O. Box 03004, E-18080 Granada, Spain, \\
  \altaffilmark{14} Weierstr 30, 8630 Rueti, Switzerland. \\
 }

\shorttitle{First study of 54 eccentric eclipsing binaries}
 \shortauthors{P. Zasche et al.}

  \date{Received \today; accepted ???}

\begin{abstract} We present an analysis of the apsidal motion and light curve parameters of 54
never-before-studied galactic Algol-type binaries. This is the first analysis of such a large
sample of eccentric eclipsing binaries in our Galaxy, and has enabled us to identify several
systems that are worthy of further study. Bringing together data from various databases and
surveys, supplemented with new observations, we have been able to trace the long-term evolution of
the eccentric orbit over durations extending back up to several decades. Our present study explores
a rather different sample of stars to those presented in the previously published catalogue of
eccentric eclipsing binaries, sampling to fainter magnitudes, covering later spectral types,
sensitive to different orbital periods with more than 50\% of our systems having periods longer
than six~days. The typical apsidal motion in the sample is rather slow (mostly of order of
centuries long), although in some cases this is less than 50~years. All of the systems, except one,
have eccentricities less than 0.5, with an average value of 0.23. Several of the stars also show
evidence for additional period variability. In particular we can identify three systems in the
sample, HD~44093, V611~Pup, and HD~313631, which likely represent relativistic apsidal rotators.
 \end{abstract}


 \section{Introduction}

\begin{table*}[h!]
\caption{Relevant information for the analysed systems.}  \label{InfoSystems}
 \scriptsize
  \centering \scalebox{0.99}{
\begin{tabular}{lccccccccccc}
   \hline\hline\noalign{\smallskip}
  System        & Other ID                 &          RA            &             DE                          &$V_{\rm max}^{\,A}$&$(J-H)^{B}$ & $(B-V)^{A}$& Sp.Type$^C$ \\
  name          &                          &                        &                                         & [mag]             &  [mag]     & [mag]&               \\
  \hline\noalign{\smallskip}                                                                                                                                              
  \object{V1137 Cas}      & GSC 04297-01664          & 01$^h$34$^m$53$^s$.930 & $+67^\circ$38$^\prime$15$^{\prime\prime}$.00 &  11\m81       &  0.255    & 0.80 &       \\          
  \object{CR Per}         & AN 16.1940               & 02$^h$09$^m$52$^s$.170 & $+57^\circ$54$^\prime$32$^{\prime\prime}$.28 &  12\m02       &  0.100    & 0.47 &     \\            
  \object{CzeV 662}       & GSC 03691-00735          & 02$^h$36$^m$48$^s$.615 & $+56^\circ$04$^\prime$11$^{\prime\prime}$.92 &  12\m96       &  0.077    & 0.36 &      \\           
  \object{CzeV 701}       & TYC 3700-608-1           & 02$^h$40$^m$23$^s$.897 & $+54^\circ$14$^\prime$34$^{\prime\prime}$.92 &  10\m81       &  0.254    & 0.52 &  \\               
  \object{CzeV 688}       & GSC 03708-01145          & 02$^h$47$^m$20$^s$.119 & $+56^\circ$15$^\prime$05$^{\prime\prime}$.63 &  13\m12       &  0.254    & 1.01 &  \\               
  \object{V1018 Cas}      & GSC 04048-00934          & 03$^h$01$^m$19$^s$.374 & $+60^\circ$34$^\prime$20$^{\prime\prime}$.24 &  10\m24       &  0.099    & 0.72 & B2III [1]  \\     
  \object{V1268 Tau}      & HIP 17168                & 03$^h$40$^m$38$^s$.770 & $+28^\circ$46$^\prime$24$^{\prime\prime}$.00 &   7\m38       &  0.050    & 0.15 & A0 [2] \\         
  \object{NO Per}         & SV$*$ SON 8551           & 04$^h$15$^m$41$^s$.820 & $+48^\circ$40$^\prime$42$^{\prime\prime}$.10 &  12\m21       &  0.216    & 0.89 &    \\             
  \object{CzeV 1279}      & UCAC4 093-005674         & 04$^h$50$^m$58$^s$.786 & $-71^\circ$31$^\prime$49$^{\prime\prime}$.35 &  11\m07       &  0.285    & 0.38 &  \\               
  \object{DT Cam}         & HIP 24390                & 05$^h$13$^m$57$^s$.689 & $+56^\circ$30$^\prime$28$^{\prime\prime}$.63 &   8\m19       &  0.054    & 0.20 & A2 [2] \\         
  \object{UCAC4 609-022916}&2MASS J05455225+3142200  & 05$^h$45$^m$52$^s$.250 & $+31^\circ$42$^\prime$20$^{\prime\prime}$.00 &  14\m83       &  0.162    & 0.52 &  \\               
  \object{V409 Cam}       & TYC 4524-1856-1          & 05$^h$46$^m$43$^s$.904 & $+75^\circ$20$^\prime$56$^{\prime\prime}$.51 &  10\m73       &  0.136    & 0.42 &  \\               
  \object{CzeV 1144}      & UCAC4 602-024605         & 05$^h$48$^m$07$^s$.920 & $+30^\circ$13$^\prime$19$^{\prime\prime}$.10 &  14\m59       &  0.158    & 0.46 &  \\               
  \object{V437 Aur}       & HIP 27469                & 05$^h$49$^m$03$^s$.060 & $+54^\circ$01$^\prime$57$^{\prime\prime}$.03 &   8\m47       & -0.016    & 0.03 & B9 [3] \\         
  \object{CzeV 364}       & GSC 02405-01470          & 05$^h$49$^m$40$^s$.700 & $+30^\circ$25$^\prime$00$^{\prime\prime}$.70 &  13\m95       &  0.236    & 0.59 &  \\               
  \object{CzeV 464}       & USNO-A2.0 1200-03882057  & 05$^h$50$^m$11$^s$.541 & $+31^\circ$19$^\prime$40$^{\prime\prime}$.34 &  15\m12       &  0.103    &      &  \\               %
  \object{TYC 3750-599-1} & GSC 03750-00599          & 05$^h$50$^m$52$^s$.961 & $+53^\circ$58$^\prime$23$^{\prime\prime}$.60 &  10\m60       & -0.072    & 0.08 &  \\               
  \object{TYC 729-1545-1} & GSC 00729-01545          & 06$^h$07$^m$18$^s$.601 & $+13^\circ$31$^\prime$45$^{\prime\prime}$.93 &   9\m26       &  0.105    & 0.29 &  \\               
  \object{CD-33 2771}     & TYC 7076-1598-1          & 06$^h$10$^m$16$^s$.479 & $-33^\circ$21$^\prime$20$^{\prime\prime}$.88 &   9\m78       &  0.724    & 1.47 & K5III+K5V [4] \\  
  \object{HD 44093}       & GSC 00140-01064          & 06$^h$20$^m$04$^s$.841 & $+04^\circ$54$^\prime$44$^{\prime\prime}$.60 &   9\m25       & -0.043    & 0.09 & B8/9V [5] \\      
  \object{TYC 5378-1590-1}& GSC 05378-01590          & 06$^h$45$^m$43$^s$.953 & $-08^\circ$50$^\prime$35$^{\prime\prime}$.50 &  10\m93       & -0.029    &-0.02 &  \\               
  \object{HD 55338}       & TYC 4823-2213-1          & 07$^h$12$^m$20$^s$.843 & $-05^\circ$25$^\prime$53$^{\prime\prime}$.94 &   9\m54       &  0.034    & 0.11 & A1IV/V [5] \\     
  \object{RW CMi}         & AN 128.1929              & 07$^h$22$^m$22$^s$.590 & $+02^\circ$26$^\prime$22$^{\prime\prime}$.30 &  12\m87       &  0.139    & 0.39 &   \\              
  \object{V611 Pup}       & HD 62589                 & 07$^h$44$^m$06$^s$.063 & $-16^\circ$55$^\prime$57$^{\prime\prime}$.95 &   8\m09       & -0.140    &-0.09 & B3III [6] \\      
  \object{CzeV 1283}      & HD 68304                 & 08$^h$09$^m$51$^s$.160 & $-45^\circ$37$^\prime$50$^{\prime\prime}$.69 &  10\m00       &  0.146    & 0.40 & F0IV/V [11] \\    
  \object{TYC 7126-2416-1}& GSC 07126-02416          & 08$^h$16$^m$30$^s$.996 & $-33^\circ$14$^\prime$38$^{\prime\prime}$.31 &  10\m37       &  0.483    & 0.93 &  \\               
  \object{CzeV 1183}      & 2MASS 08283756-4351041   & 08$^h$28$^m$37$^s$.562 & $-43^\circ$51$^\prime$04$^{\prime\prime}$.11 &  12\m43       &  0.251    & 0.77 &  \\               
  \object{DK Pyx}         & HIP 41980                & 08$^h$33$^m$24$^s$.062 & $-34^\circ$38$^\prime$55$^{\prime\prime}$.31 &   7\m84       & -0.114    &-0.10 & B3III [7] \\      
  \object{PS UMa}         & TYC 4375-1733-1          & 08$^h$56$^m$46$^s$.479 & $+69^\circ$40$^\prime$32$^{\prime\prime}$.12 &  12\m45       &  0.311    & 0.54 &   \\              
  \object{HD 87803}       & HIP 49354                & 10$^h$04$^m$31$^s$.513 & $-69^\circ$21$^\prime$20$^{\prime\prime}$.27 &   9\m51       &  0.013    & 0.06 & B9.5V [8] \\      
  \object{TYC 8603-723-1} & GSC 08603-00723          & 10$^h$06$^m$25$^s$.330 & $-55^\circ$00$^\prime$44$^{\prime\prime}$.60 &  11\m38       & -0.027    & 0.08 &  \\               
  \object{HD 306001}      & GSC 08958-03048          & 11$^h$06$^m$07$^s$.882 & $-61^\circ$09$^\prime$09$^{\prime\prime}$.25 &   9\m58       & -0.071    &-0.09 & B5 [9] \\         
  \object{TYC 8217-789-1} & GSC 08217-00789          & 11$^h$15$^m$06$^s$.983 & $-48^\circ$15$^\prime$33$^{\prime\prime}$.44 &  11\m28       &  0.186    & 0.65 &  \\               
  \object{TYC 9432-1633-1}& 2MASS J14444107-7721530  & 14$^h$44$^m$41$^s$.077 & $-77^\circ$21$^\prime$53$^{\prime\prime}$.02 &  11\m70       &  0.290    & 0.69 &  \\               
  \object{SS TrA}         & GSC 09022-01335          & 15$^h$39$^m$21$^s$.650 & $-60^\circ$53$^\prime$38$^{\prime\prime}$.30 &  11\m02       &  0.106    & 0.27 &  \\               
  \object{KO Nor}         & TYC 8719-163-1           & 16$^h$25$^m$59$^s$.508 & $-56^\circ$55$^\prime$03$^{\prime\prime}$.97 &  10\m85       &  0.082    & 0.62 &  \\               
  \object{V883 Sco}       & HD 152901                & 16$^h$57$^m$52$^s$.443 & $-37^\circ$59$^\prime$47$^{\prime\prime}$.57 &   7\m04       &  0.013    & 0.05 &  B2V [10] \\      
  \object{V1301 Sco}      & GSC 07368-01457          & 17$^h$05$^m$18$^s$.635 & $-34^\circ$56$^\prime$00$^{\prime\prime}$.71 &  13\m03       &  0.228    & 0.68 &  \\               
  \object{HD 158801}      & TYC 7896-1604-1          & 17$^h$33$^m$19$^s$.034 & $-43^\circ$15$^\prime$01$^{\prime\prime}$.40 &   9\m54       &  0.060    & 0.27 & A5/7II [11] \\    
  \object{TYC 6258-1011-1}& 2MASS J17533294-2031094  & 17$^h$53$^m$32$^s$.945 & $-20^\circ$31$^\prime$09$^{\prime\prime}$.53 &  12\m08       &  0.182    & 0.48 &   \\              
  \object{HD 163735}      & TYC 5095-296-1           & 17$^h$57$^m$54$^s$.027 & $-05^\circ$41$^\prime$10$^{\prime\prime}$.15 &   9\m62       &  0.251    & 0.54 &  F3V [5] \\       
  \object{HD 313631}      & TYC 6842-1455-1          & 18$^h$00$^m$10$^s$.195 & $-23^\circ$53$^\prime$46$^{\prime\prime}$.10 &  10\m41       &  0.170    & 0.58 & OB [12] \\        
  \object{HD 164610}      & GSC 07899-00130          & 18$^h$03$^m$39$^s$.099 & $-37^\circ$43$^\prime$47$^{\prime\prime}$.68 &   8\m64       &  0.052    & 0.22 &  A1mA8-A8 [7]  \\ 
  \object{V1344 Her}      & HD 348698                & 18$^h$27$^m$18$^s$.443 & $+19^\circ$08$^\prime$33$^{\prime\prime}$.32 &  11\m69       &  0.228    & 0.27 & G0 [9]  \\        
  \object{HD 170749}      & TYC 7398-2681-1          & 18$^h$32$^m$34$^s$.765 & $-33^\circ$14$^\prime$40$^{\prime\prime}$.85 &   9\m94       &  0.060    & 0.17 & A0/1V [7] \\      
  \object{TYC 8378-252-1} & GSC 08378-00252          & 18$^h$59$^m$51$^s$.282 & $-47^\circ$11$^\prime$47$^{\prime\prime}$.75 &  11\m03       &  0.168    & 0.58 &   \\              
  \object{TYC 6303-308-1} & 2MASS J19393409-1739553  & 19$^h$39$^m$34$^s$.096 & $-17^\circ$39$^\prime$55$^{\prime\prime}$.45 &  11\m31       &  0.110    & 0.37 &   \\              
  \object{PS Vul}         & HIP 9709                 & 19$^h$43$^m$55$^s$.974 & $+27^\circ$08$^\prime$07$^{\prime\prime}$.43 &   6\m46       &  0.768    & 1.02 &  K3II+B6 [13] \\  
  \object{V839 Cep}       & TYC 3964-741-1           & 21$^h$03$^m$31$^s$.714 & $+59^\circ$25$^\prime$50$^{\prime\prime}$.41 &   9\m73       &  0.055    & 0.36 &  B8 [14] \\       
  \object{TYC 5195-11-1}  & 2MASS J21264316-0031104  & 21$^h$26$^m$43$^s$.166 & $-00^\circ$31$^\prime$10$^{\prime\prime}$.53 &  11\m32       &  0.233    & 0.49 &   \\              
  \object{TYC 2712-1201-1}& GSC 2712-1201            & 21$^h$34$^m$57$^s$.620 & $+35^\circ$12$^\prime$51$^{\prime\prime}$.46 &  10\m70       & -0.015    &-0.01 &   \\              
  \object{UCAC4 585-123180}& NSVS 8774343            & 21$^h$39$^m$21$^s$.030 & $+26^\circ$52$^\prime$36$^{\prime\prime}$.70 &  12\m88       &  0.264    & 0.57 &   \\              
  \object{V922 Cep}       & TYC 4481-230-1           & 23$^h$01$^m$39$^s$.222 & $+69^\circ$42$^\prime$44$^{\prime\prime}$.96 &  11\m39       &  0.137    & 0.29 &   \\              
  \object{V389 And}       & HIP 116153               & 23$^h$32$^m$01$^s$.312 & $+43^\circ$49$^\prime$20$^{\prime\prime}$.49 &   8\m56       & -0.031    & 0.19 &  A0 [15]  \\      
 \noalign{\smallskip}\hline
\end{tabular}}\\
\scriptsize Note: [A] - value based on APASS \citep{2015AAS...22533616H} or Tycho catalogue
\cite{2000A&A...355L..27H}, [B] - 2MASS catalogue, \cite{2006AJ....131.1163S}, [C] - Various
published papers: [1] - \cite{1978A&AS...32...25R}, [2] - \cite{1983RGOB..189....1T}, [3] -
\cite{1933PUSNO..13....1M}, [4] - \cite{2009MNRAS.395..593P}, [5] - \cite{1999MSS...C05....0H}, [6]
- \cite{1988mcts.book.....H}, [7] - \cite{1982mcts.book.....H}, [8] - \cite{1975mcts.book.....H},
[9] - \cite{1995A&AS..110..367N}, [10] - \cite{1977ApJS...35..111G}, [11] -
\cite{1978mcts.book.....H}, [12] - \cite{1971PW&SO...1a...1S}, [13] - \cite{2002ApJS..143..513G},
[14] - \cite{1958TrRig...7...33A}, [15] - \cite{1935sgcs.book.....D}.
\end{table*}

For independent testing of general relativity, as well in confronting stellar structure models,
eccentric eclipsing binary systems are often mentioned as being ideal astrophysical laboratories.
However, in the last few decades the exploitation of eccentric eclipsing binaries (hereafter EEBs)
to current astrophysical research problems has been developed, thanks to the large photometric
surveys and availability of precise data, including for fainter targets. Studies have focused on,
in particular, the period-eccentricity distribution of the binary systems, which provides a crucial
test of our models of star formation, theory of orbital circularization, and dynamical evolution of
binaries and multiple systems, amongst others. These studies have asked questions such as: Is the
formation of binary and multiple star systems solely a matter of fragmentation, accretion, or some
N-body dynamics, or is it a fruitful combination of all these mechanisms? Is the subsequent orbit
migration crucial for explaining the observed orbital properties of binaries and multiples? The
distribution of eccentricity can provide strong constraints on star forming theories and can be
compared with stellar formation models and N-body simulations (see e.g.
\citealt{2008MNRAS.389..925T}, or \citealt{2012ApJ...751....4K}). The number of known systems
suitable for detailed modelling is still rather limited, and studies like the present one
significantly extend the available sample and help us to answer some of the questions above.

Study of the light curves (hereafter LCs) of eclipsing binaries and their modelling with available
tools (i.e. a well-known Wilson-Devinney algorithm, see \citealt{1971ApJ...166..605W} and
\citealt{1979ApJ...234.1054W}) can be used for deriving the value of orbital eccentricity. On the
other hand, the orbital period is mostly a known quantity. Moreover, the LC modelling also provides
information about the physical properties of both components, their relative luminosities or
relative radii (with respect to the semimajor axis). Additionally, the results of such an analysis
can help to estimate the internal structure constants for the particular system.

The investigation of period changes in EEB systems on the basis of their minima timings variation
(both primary and secondary) alone is a familiar method in stellar astrophysics as described
in several seminal papers, for example, \cite{1983Ap&SS..92..203G}, or \cite{1995Ap&SS.226...99G}. As a
very short description of the method, the sidereal and anomalistic periods of the binary are
connected with the relation $P_s = P_a (1-\dot \omega /{2\pi}),$ where $\dot \omega$ is the rate of
the apsidal motion $\omega = \omega_0 + \dot \omega E.$ The period of such revolution is then $U =
2\pi P_a/\dot\omega.$ The individual equations for the computation of the time of primary and
secondary minima were published, for example in \cite{1983Ap&SS..92..203G}. We also suppose that the
apsidal motion remains linear in time.

The number of new photometric observations of eclipsing binaries is increasing rapidly
year-on-year. This is partly due to non-professional astronomers, but mostly because of new
photometric surveys covering the whole sky. Using these data, it is possible to derive many new
times of eclipses at various time epochs, and to trace periodic changes in these systems, that is,
detect the third bodies, or study the apsidal motion in them. Examples of such studies have been
published quite frequently during the last few years, see for example \cite{2007MNRAS.378..757P},
\cite{2014AJ....147..130Z}, or \cite{2016MNRAS.455.4136B}.

However, despite increasing the number of data points, a detailed analysis
is still lacking for some of the systems
. Hence, for our analysis we have used observations from automated photometric
surveys, from satellites, from our new ground-based observations, as well as from previously
published data from various publications.

The advantage of our method for deriving the eccentricity is obvious. Without having any
information about the masses (no spectroscopy and RV solutions available for these systems), our
solution still has to be considered preliminary. However, having an estimate of the eccentricity
from the LC solution, it becomes possible to perform a period analysis of the eclipse times to
refine its value, and vice versa. Sometimes the LC coverage is insufficient to derive a reliable
value of the eccentricity, but the $O-C$ diagram with times of eclipses can help us. Sometimes the
change of $\omega$ is so slow that a trustworthy analysis cannot be done using only minima times,
but the LC solution can reveal the eccentricity more reliably.

For our target selection the selected systems should not have been studied before, meaning that no
LC or period analysis has been published. Although some of the systems have had their orbital
periods mentioned in previous publications, no further information or analysis was available. All
the systems are well detached (as needed for an eccentric-orbit binary), located in both the
southern and northern hemispheres, and their respective orbital periods range from several days to
several dozen days. The only input parameters needed for the analysis are the period and an
estimate of the primary temperature (see below). In the whole study, we see rather heterogeneous
effort from the various observers (mainly professionals, but sometimes also non-professionals) and
observational strategies (sometimes dedicated observations, sometimes only by chance products of
observing another target in the field). Some of the selected stars are relatively faint, although
several of these systems are sufficiently bright (mag $<$ 10) that they were observed by the
Hipparcos satellite \citep{1997ESASP1200.....E}. We consider the present paper as a starting point
for some future, more focused, observational effort, or to aid the target selection for a future
dedicated studies of these targets.

An overview of the basic parameters of the analysed systems is presented in Table
\ref{InfoSystems}, where the individual values are taken from the available databases and
catalogues (SIMBAD, GCVS - \citealt{2017ARep...61...80S}). The primary temperature needed for the
LC analysis was in most cases estimated from the photometric indices, but for those systems for
which a spectral classification has been published, this was used for a temperature estimation of
the fixed $T_1$ value used for the LC analysis. For most systems, the GCVS designation is still
missing, hence we used some most common catalogue information taken from CDS/SIMBAD, together with
the precise J2000.0 coordinates for better identification of the star.

For analysis of the LC as well as deriving new minima times, several different sources and
databases were used together. These data, along with our new observations, present material that is
sometimes very suitably complementary. For example, sometimes the phase coverage is poor in one
source, but it may have good photometry in another, and vice versa. See our final plots below in
Figures \ref{FigLCOC1}, and \ref{FigLCOC2} to \ref{FigLCOC9}.

\begin{figure*}[h]
  \centering
  \includegraphics[width=0.92\textwidth]{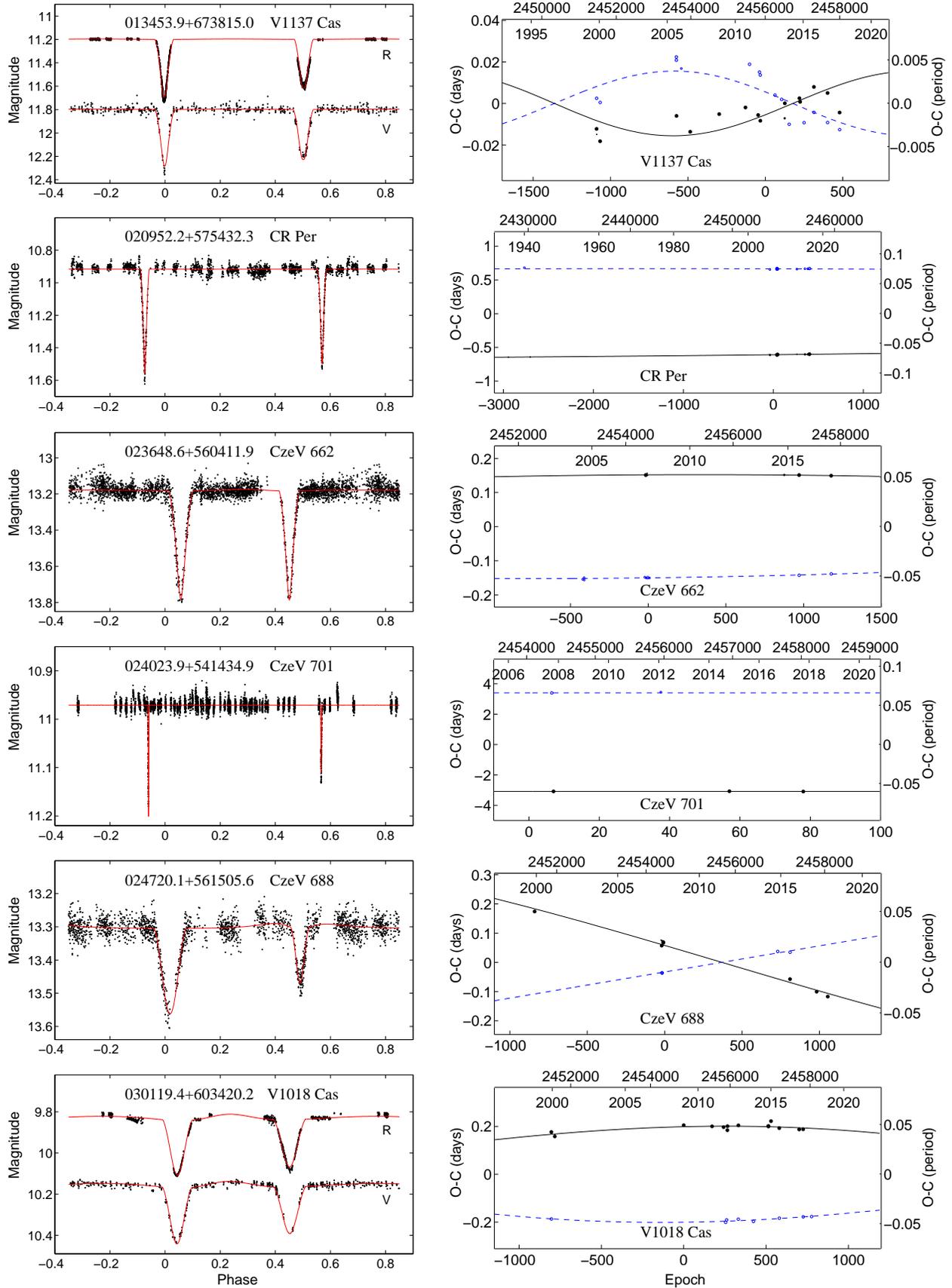}
  \caption{Plot of the light curves and $O-C$ diagrams of the analysed systems. The small letters denote the individual filters
  for the particular light curve (standard notation, while "C" stands for unfiltered data and "S" indicates
  the special SuperWASP filter). For the $O-C$ diagrams the full dots stand for the primary minima (as well as
  the solid line), while the open circles represent the secondary minima (and the dashed curve).}
  \label{FigLCOC1}
\end{figure*}

\section{Approach for the analysis}

Despite the fact that several of the systems have rather low photometric
variation amplitudes, and some are also relatively faint targets, usually some of the automatic photometric
surveys can provide us with reliable photometry to facilitate LC modelling. Hence, when having
different sources of data from different surveys, databases, and publications, we used the best of these
for the LC solution (i.e. those with the lowest scatter and the best phase coverage of the complete
light curve).

For the light-curve analysis we used the code {\sc PHOEBE} \citep{2005ApJ...628..426P}, which is
originally based on the Wilson-Devinney algorithm (\citealt{1971ApJ...166..605W} and
\citealt{1979ApJ...234.1054W}) and its later modifications. For those systems where photometry in
different filters was used, we analysed these data simultaneously.

Because spectroscopic studies with radial velocities are missing for all these systems, there are
several assumptions which have to be considered. At first, the mass ratio of all the systems was
kept at a value of unity. This approach is justified because all the systems are well-detached and
the ellipsoidal variations outside of their minima are almost negligible. For such systems the
photometric mass ratio can only be approximately estimated, as quoted for example by
\citealt{2005ApSS.296..221T}. Other physical parameters were instead estimated and derived in a
relative sense, rather than in absolute units (e.g. radii, luminosities), hence these values are
still only approximate estimates and should not be used as fundamental parameter sources.

Due to all these reasons we used the following approach for the analysis. Firstly, a very first
rough LC analysis was performed. Second, the initial LC analysis was used to estimate the available
minima, which were then analysed to estimate preliminary apsidal motion parameters (with the
assumption $i=90^\circ$). Third, the eccentricity ($e$), argument of periastron ($\omega$) and
apsidal motion rate ($\dot \omega$) resulting from the apsidal motion analysis were used for the
preliminary light curve analysis. Fourth, a further parameter from the LC analysis, the inclination
($i$), was then used for the apsidal motion analysis. And finally, the resulting $e$, $\omega$, and
$\dot \omega$ values from the apsidal motion analysis were used for the final LC analysis.

The AFP (automatic fitting procedure) method for deriving the individual times of minima was the
same as presented in \cite{2014A&A...572A..71Z}, applied to different photometric databases. All the minima times used for the analysis were given in Table \ref{MINIMA}.

For all of the systems, the presence of the third light was tested. This new free parameter was
only applied during the last step of the fitting process, because it is a very sensitive
second-order light curve parameter, which could only be conclusively detected in high quality LCs.

\section{Results}

The selection of targets for the present analysis was straightforward, and did not depend on where
the star is located, nor how bright or faint it is. The criterion was based only whether the star
had been previously studied, as our main aim was to enlarge the existing set of eclipsing binaries
in the period - eccentricity diagram.

Another selection criterion was the data coverage of the particular binary. Only those systems with
well-sampled phase light curve were included into our sample. This means that photometry should
exist and especially that it should be adequately precise near both eclipses, so that their
duration and depths could be recovered. Sometimes we had to use more data sources for this
analysis (because in one dataset the coverage of minima is only poor).

The stars included in our sample can be divided into three groups according to their discoveries of
photometric periodicity. Some have already been mentioned in the former ASAS database
\citep{2002AcA....52..397P}, a second group were serendipitous discoveries already mentioned in the
literature (mainly the so-called CzeV catalogue, \citealt{2017OEJV..185....1S}), and finally the
third group consists of stars which have not yet been reported as eclipsing binaries, and this is
for the first time that such stars have been identified as being an EEB system.

For the analysis we usually used some of the following available databases and surveys: ASAS
\citep{2002AcA....52..397P}, SuperWASP \citep{2006PASP..118.1407P}, NSVS
\citep{2004AJ....127.2436W}, ASAS-SN (\citealt{2014ApJ...788...48S} and
\citealt{2017PASP..129j4502K}), TAROT \citep{2001A&A...378...76B}, INTEGRAL-OMC
\citep{2003A&A...411L.261M}, Bochum Galactic Disk Survey \citep{2015AN....336..590H}, CRTS
\citep{2009ApJ...696..870D}, and sometimes also Hipparcos \citep{1997A&A...323L..49P} when
available for brighter sources. For one system (V1268~Tau) the photometry from the STEREO satellite
(\citealt{2008SSRv..136....5K} and \citealt{2011MNRAS.416.2477W}) was also used. Our new photometry
was obtained with various telescopes at different observatories (BOOTES in Spain,
FRAM\footnote{FRAM \citep{2014RMxAC..45..114E} telescope is part of the Pierre Auger Observatory
\citep{PierreAuger}.} in Argentina, Danish 1.54-m telescope in Chile, 65-cm telescope in
Ond\v{r}ejov, Czech Republic), but also with a non-negligible contribution by several amateur
observers with their small telescopes. For some of the systems we collected the data for more than
ten~years. The already known stars with published minima timings were checked and the data
downloaded from the $O-C$ gateway\footnote{See http://var2.astro.cz/ocgate/}
\citep{2006OEJV...23...13P}.

The crucial parameters resulting from our light curve and apsidal motion fits are listed in Table
\ref{LCOCparam}. For those stars where the apsidal motion is very slow, we have only indicated that
the apsidal period is very long ($>1000$yr), and that our data coverage is still too poor to
facilitate a reliable analysis. This can clearly be seen in the Figs. \ref{FigLCOC1}, and
\ref{FigLCOC2} to \ref{FigLCOC9} with the $O-C$ diagrams. The parameters $HJD_0$ and $P$ stand for
the average ephemerides representing the value $O-C=0$ in the $O-C$ diagrams, meaning that these
are not suitable for planning the observations.

Concerning the presented solutions given in Table \ref{LCOCparam}, we still have to emphasize that
these are still only the preliminary results, based on photometry only. The same applies for the
errors of individual parameters given in Table \ref{LCOCparam}, which are based on the errors given
by the PHOEBE program. Sometimes these errors are rather underestimated.

What definitely should be mentioned is that several systems have quite unrealistic parameters
concerning their temperatures, radii, and luminosities. We are aware of the fact that the solution
presented here (having q\,=\,1.0) is unlikely, or at least improbable. Hence, for these problematic
cases we decided to use the approach for deriving the mass ratio from the luminosity ratio given in
\cite{2003MNRAS.342.1334G}. This method uses the assumption that both components are located on the
main sequence. However, using this approach we found out that for a few such cases even this method
is not able to describe the observed data adequately and provide us with physically unreliable
result. These are the cases like V1137~Cas, or CzeV~1279. It may be caused by the fact that
the mass ratio derivation is based on the assumption of the main sequence components. For some
others (such as e.g. NO~Per, CzeV~364, SS~TrA, V1301~Sco, V1344~her, or PS~Vul) the components are
probably giants, or some overluminous stars according to their effective temperatures. Only further,
more detailed, study would be able to reveal their true nature.
Because the number of systems is rather large, we cannot focus on all of the systems in detail,
hence we would like to point out here only the most interesting targets.

\begin{table}[t]
 \caption{Heliocentric minima of the systems used for our analysis.}
 \label{MINIMA}
 \scriptsize
 \centering
 \begin{tabular}{l l l c c l}
 \hline\hline
 Star  & HJD - 2400000 & Error  & Type & Filter & Reference\\
 \hline
 V1137 Cas & 51473.568   &        & P & V & IBVS 5570 \\
 V1137 Cas & 53613.4050  & 0.0002 & S & R & IBVS 5741 \\
 V1137 Cas & 53613.40344 & 0.0002 & S & R & OEJV 74 \\
 V1137 Cas & 53615.45611 & 0.0003 & P & R & New - This study \\
 V1137 Cas & 53746.48711 & 0.0009 & S & R & New - This study \\
 \dots \\
 \hline \hline
\end{tabular}
 \begin{list}{}{}
 \item Note: Table is published in its entirety in the electronic supplement of the journal, at CDS and
 also below in the appendix section at the end of the manuscript.
 \end{list}
\end{table}

Amongst the systems reported in this study, several systems were found to contain additional close
visual components. These are: V1018~Cas, V1268~Tau, HD~55338, V611~Pup, V839~Cep, V922~Cep, and
V389~And, respectively. For some of them a non-zero value of the third light was also detected
during the light curve solution.

\begin{figure*}
  \centering
  \includegraphics[width=0.950\textwidth]{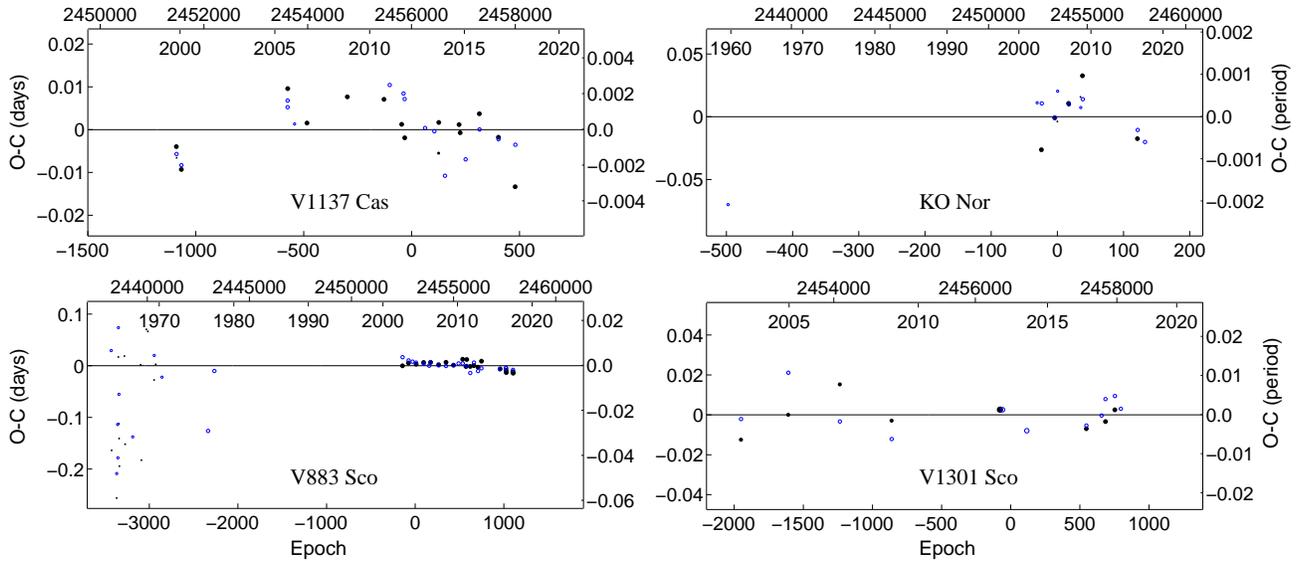}
  \caption{O-C diagrams after subtraction of the apsidal motion term. These four systems are
  suspected for some additional variation of their minima times, hence deserve a special attention.}
  \label{OCresid}
\end{figure*}

One system is a giant star, namely CD-33~2771, which merits special attention for future
investigation. It has also the longest orbital period of about 200~days. Because this star has such
a long period it would not be expected to show wide eclipses when it is not a giant star. According
to our modelling, the primary component is probably of luminosity class III, while the secondary of
subgiant class IV.

Also remarkable are systems where some possible additional variation in the $O-C$ diagram appears
after subtraction of the apsidal motion term. These are mainly: V1137~Cas, KO~Nor, V883~Sco, and
V1301~Sco. These should be suitable for further observations to detect these changes, and to
confirm the third-body hypothesis or other phenomena. See Fig. \ref{OCresid} for their $O-C$
diagrams after subtraction of the apsidal motion.

Amongst all of the studied systems, we have chosen those with given some spectral type estimates
and derived the relativistic contribution to the total apsidal motion. This is viable only when we
know the masses, in cases when the temperatures were only derived using the photometric indices the
masses are too uncertain for any such analysis. Following the method by \cite{1985ApJ...297..405G}
we derived the relativistic apsidal advance and for those systems with the largest contribution we
gave their values in Table \ref{TabRelativistic}. As one can see, these values are affected by
large errors due to the fact that the apsidal advance is very slow and only small fraction of the
whole apsidal period is covered with data nowadays. These systems also deserve detailed study in
the future.

Eccentric eclipsing binaries which have been studied quite frequently, and the known systems were
also included in various catalogues of EEBs. The first one is for example, by \cite{1988BICDS..35...15H},
later updated by \cite{1999AJ....117..587P}, while the most recent compilation of those stars is
presented in \cite{2007MNRAS.378..179B}. Their catalogue contains a total of 108 systems with known
eccentricities and apsidal motion periods. Therefore, our set of 54 new EEBs presents a significant
contribution to the topic.

\begin{figure}
  \centering
  \includegraphics[width=0.4\textwidth]{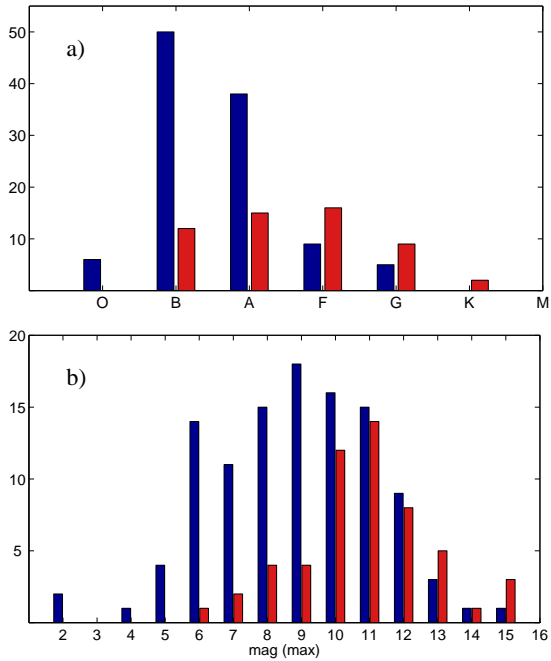}
  \caption{Comparison of our sample of eccentric binaries (red) with the catalogue of eccentric
  systems by \cite{2007MNRAS.378..179B} in blue. Upper plot: Distribution of spectral types. Lower plot:
  Apparent magnitudes of the systems.}
  \label{Statistics}
\end{figure}

One can compare these two sets of stars and plot several interesting statistics. As can be seen in
Fig. \ref{Statistics}, our new systems have slightly later spectral types, and are also rather
fainter than the original sample by \cite{2007MNRAS.378..179B}. This is caused by the fact that we
mainly used the photometry from the various databases and automatic surveys, which are typically
focused on stars fainter than 10 mag. On the other hand, the catalogue by
\cite{2007MNRAS.378..179B} presented all the well-known systems that have been studied for up to
decades or even more, and that are sometimes as bright as naked-eye stars. The distribution of
spectral types is a little more complicated to describe due to the fact that we do not know the
interstellar reddening for most of these stars, and that the assumed spectral types were only
derived on the basis of their photometric indices (i.e. a kind of lower limit). However, the stars
having early spectral types have mostly been studied during the last century, and it is only during
the last decades with the use of large telescopes did the focus of stellar astronomers shift to the
more late and fainter systems. We also have more systems with longer orbital periods ($>30$~d) than
appear in \cite{2007MNRAS.378..179B}.

\begin{figure}
  \centering
  \includegraphics[width=0.43\textwidth]{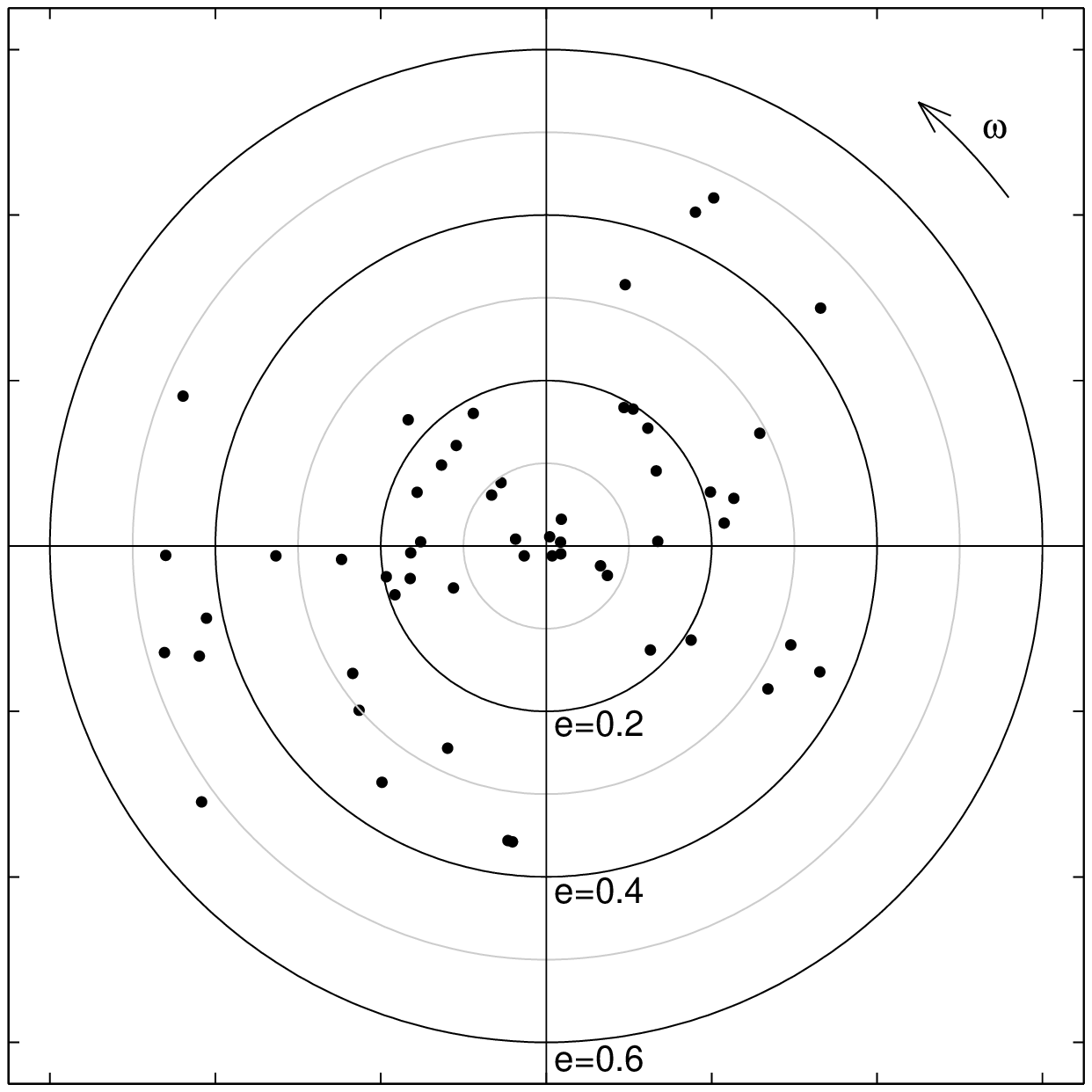}
  \caption{Distribution of omega angles with respect to the eccentricity.}
  \label{Omeg-e}
\end{figure}

We can also address the issue of selection effects for our sample of stars. These "serendipitous"
discoveries should be more likely detected as eccentric when the two minima are displaced from
their positions 0.0 and 0.5 in phase diagram, but the binaries which have $\omega$ close to
90$^\circ$ or 270$^\circ$ should be only rarely detected. This can more easily be done when the
photometry is sufficiently precise and also when the eccentricity is higher (i.e. the durations of
both eclipses are significantly different). Hence, we tried to test this distribution of $\omega$
versus eccentricity in our Fig. \ref{Omeg-e}. One can see that the diagram is almost uniformly
covered with data points, but the number of systems with higher eccentricity is still rather low
(see Sect. 4 below).

\begin{table*}
 \caption{The parameters of the light curve fits and the apsidal motion.}
 \label{LCOCparam}
 \footnotesize
 \centering \scalebox{0.79}{
 \begin{tabular}{l c r c c c c c c c r c c c c c c c}
 \hline\hline
  System         &  $i$       & \multicolumn{1}{c}{$T_1$} & \multicolumn{1}{c}{$T_2$} &  $L_1$  &  $L_2$  & $L_3$  & $R_1/a$ & $R_2/a$ & $HJD_0     $ & \multicolumn{1}{c}{$P$ [d]} &  $e$   & $\omega$ [deg] & $U$ [yr] \\
                 &  [deg]     & \multicolumn{1}{c}{[K]}   & \multicolumn{1}{c}{ [K]}  &  [\%]   &   [\%]  &  [\% ] &         &         & $[2400000+]$ &            &        &                &          \\
 \hline
 V1137 Cas       & 88.19(0.45)&  6000 &  5938(80) &  68.5(0.9)&  31.5(0.6)&  0.0     & 0.130(6)&  0.090(7)& 56002.7216   &  4.1589888 & 0.012(0.017) &  69.2 (12.8) & 35.1 (7.3) \\
 CR Per          & 89.25(0.22)& 20000 & 18995(427)&  52.0(1.9)&  48.0(2.0)&  0.0     & 0.051(3)&  0.051(2)& 54005.9671   &  8.8096975 & 0.234(0.016) &  14.2 (8.2)  & $>$ 1000   \\
 CzeV 662        & 87.68(0.24)&  7500 &  7474(110)&  51.9(0.8)&  48.1(0.7)&  0.0     & 0.135(3)&  0.131(2)& 54417.7801   &  2.8943665 & 0.164(0.033) & 182.9 (3.7)  & 232.7 (59) \\
 CzeV 701        & 89.27(0.11)&  6500 &  6258(72) &  53.8(0.7)&  46.2(0.9)&  0.0     & 0.011(2)&  0.011(2)& 54032.0336   & 51.2568523 & 0.209(0.012) &  18.2 (0.5)  & $>$ 1000   \\
 CzeV 688        & 81.77(0.28)&  6000 &  4652(126)&  84.0(0.6)&  16.0(0.5)&  0.0     & 0.168(3)&  0.154(3)& 54417.3197   &  3.4676851 & 0.360(0.032) & 263.5 (0.7)  & 193.3 (12) \\
 V1018 Cas       & 79.69(0.21)& 20000 & 17912(260)&  50.8(0.5)&  49.2(0.5)&  0.0     & 0.234(2)&  0.163(8)& 54833.4983   &  4.1277539 & 0.152(0.021) & 178.1 (7.9)  & 126.1 (5)  \\
 V1268 Tau       & 86.99(0.06)& 10000 &  7166(20) &  78.3(0.3)&  19.5(0.3)&  2.2(0.6)& 0.069(2)&  0.063(2)& 56205.8772   &  8.1612083 & 0.301(0.002) & 221.3 (0.5)  & $>$ 1000   \\
 NO Per          & 89.93(0.23)&  6200 &  6169(95) &  53.2(1.8)&  23.7(0.7)& 23.1(2.2)& 0.169(2)&  0.115(2)& 57048.6114   &  5.6923047 & 0.280(0.007) & 213.4 (0.8)  & $>$ 1000   \\
 CzeV 1279       & 89.29(0.17)&  5750 &  5815(117)&  18.8(1.7)&  81.2(2.6)&  0.0     & 0.015(3)&  0.030(2)& 57017.4806   & 28.8680283 & 0.420(0.008) & 192.0 (0.9)  & $>$ 1000   \\
 DT Cam          & 87.69(0.19)&  8800 &  7382(75) &  66.4(0.7)&  33.6(0.6)&  0.0     & 0.068(3)&  0.064(2)& 56203.8349   &  7.0662440 & 0.188(0.010) &  49.2 (1.3)  & $>$ 1000   \\
 UCAC4 609-022916& 81.58(0.22)&  6600 &  6535(59) &  59.9(2.1)&  40.1(2.3)&  0.0     & 0.213(6)&  0.177(6)& 57775.5551   &  1.6736231 & 0.161(0.016) &  34.3 (1.2)  & 64.6 (19)  \\
 V409 Cam        & 84.40(0.32)&  6500 &  6585(128)&  51.4(1.1)&  47.1(1.5)&  1.5(0.6)& 0.099(2)&  0.094(2)& 53123.6893   &  6.6764702 & 0.038(0.022) & 167.6 (18.5) & 158.9 (37) \\
 CzeV 1144       & 78.14(0.17)&  6600 &  6321(73) &  58.3(0.6)&  41.7(0.6)&  0.0     & 0.173(3)&  0.163(2)& 57752.3849   &  2.0833922 & 0.192(0.015) &  60.7 (0.8)  & 298.5 (99) \\
 V437 Aur        & 88.16(0.34)& 11000 & 10981(48) &  62.4(0.8)&  37.6(0.8)&  0.0     & 0.090(4)&  0.065(3)& 56338.8586   & 11.7938027 & 0.226(0.007) & 137.6 (4.2)  & $>$ 1000   \\
 CzeV 364        & 79.37(0.30)&  6000 &  5186(81) &  55.5(0.7)&  40.6(2.2)&  3.9(1.8)& 0.137(5)&  0.158(4)& 51522.9781   &  2.5522095 & 0.364(0.008) & 335.3 (1.7)  & $>$ 1000   \\
 CzeV 464        & 88.20(0.21)&  7200 &  7075(69) &  52.1(0.4)&  47.9(0.4)&  0.0     & 0.050(4)&  0.050(4)& 56408.9827   & 10.4774146 & 0.018(0.014) &  14.6 (25.2) & 835.8 (690)\\
 TYC 3750-599-1  & 77.80(0.17)& 13000 & 12730(78) &  53.2(0.6)&  46.1(1.0)&  0.7(0.5)& 0.210(4)&  0.202(3)& 56280.6576   &  2.0597637 & 0.090(0.011) & 137.0 (2.1)  & 68.6 (4.7) \\
 TYC 729-1545-1  & 74.92(0.38)& 10000 & 10940(212)&  37.7(1.3)&  62.3(1.0)&  0.0     & 0.183(5)&  0.216(5)& 53504.8365   &  5.0778902 & 0.217(0.031) &   7.3 (12.5) & 143.7 (120)\\
 CD-33 2771      & 89.43(0.82)&  4400 &  3943(99) &  86.2(1.8)&  11.3(0.7)&  2.5(1.9)& 0.191(8)&  0.105(7)& 52327.9925   &199.9666008 & 0.123(0.079) & 204.3 (18.7) & $>$ 1000   \\
 HD 44093        & 85.27(0.32)& 11000 & 11434(157)&  39.7(2.1)&  60.3(1.9)&  0.0     & 0.133(7)&  0.149(8)& 52637.0168   &  5.9414804 & 0.442(0.078) &  65.9 (11.0) & $>$ 1000   \\
 TYC 5378-1590-1 & 89.85(0.16)& 10000 &  9380(201)&  64.1(0.8)&  35.9(0.8)&  0.0     & 0.123(2)&  0.098(3)& 53946.4904   &  3.7323527 & 0.319(0.023) & 338.0 (6.8)  & $>$ 1000   \\
 HD 55338        & 87.20(2.45)&  9200 &  8591(49) &  15.7(2.0)&  11.1(1.1)& 73.2(2.3)& 0.264(9)& 0.245(15)& 53023.7640   &  1.2114599 & 0.014(0.007) & 300.6 (2.3)  & 31.3 (3.1) \\
 RW CMi          & 86.74(0.27)&  6700 &  6870(96) &  38.8(0.9)&  61.2(0.9)&  0.0     & 0.076(3)&  0.090(2)& 57046.9677   &  6.0838012 & 0.479(0.065) & 195.6 (22.3) & $>$ 1000   \\
 V611 Pup        & 80.96(0.24)& 17000 & 19859(179)&  39.6(0.7)&  60.4(0.7)&  0.0     & 0.142(2)&  0.153(2)& 53500.8872   &  6.3178011 & 0.160(0.009) & 142.3 (11.4) & $>$ 1000   \\
 CzeV 1283       & 89.50(0.29)&  7200 &  7199(88) &  40.9(1.0)&  45.0(1.1)& 14.1(5.2)& 0.032(4)&  0.034(3)& 57040.2488   & 39.9810718 & 0.460(0.007) & 181.4 (8.7)  & $>$ 1000   \\
 TYC 7126-2416-1 & 87.17(0.30)&  4800 &  4280(132)&  66.8(2.7)&  33.2(2.6)&  0.0     & 0.048(2)&  0.048(3)& 53531.8932   &189.2707501 & 0.192(0.009) & 197.9 (9.2)  & $>$ 1000   \\
 CzeV 1183       & 87.43(0.21)&  6000 &  5799(77) &  55.6(3.2)&  31.3(2.8)& 13.1(2.4)& 0.210(9)&  0.172(8)& 53497.1451   &  2.5814597 & 0.209(0.014) & 327.0 (3.3)  & 83.8 (12)  \\
 DK Pyx          & 78.41(0.66)& 17000 & 14549(190)&  81.4(2.3)&  17.1(1.6)&  1.5(1.4)& 0.159(5)&  0.085(7)& 53123.0475   &  6.1784209 & 0.327(0.011) & 182.1 (6.9)  & 400.7 (97) \\
 PS UMa          & 85.39(0.38)&  6500 &  6535(205)&  53.4(6.8)&  46.6(5.2)&  0.0     & 0.094(4)&  0.086(8)& 51507.9642   &  9.2716673 & 0.094(0.022) & 125.6 (13.7) & $>$ 1000   \\
 HD 87803        & 87.25(0.09)& 10400 & 10386(72) &  50.2(1.3)&  45.9(1.0)&  3.9(1.2)& 0.053(3)&  0.051(2)& 53511.1799   & 11.5105711 & 0.475(0.010) & 157.6 (10.0) & $>$ 1000   \\
 TYC 8603-723-1  & 84.78(0.40)&  9500 &  8032(111)&  68.1(1.7)&  26.3(5.0)&  5.6(3.9)& 0.121(5)&  0.101(4)& 53511.7334   &  4.4111698 & 0.439(0.031) &  40.9 (8.4)  & $>$ 1000   \\
 HD 306001       & 78.65(0.78)& 15000 & 13852(215)&  54.2(4.0)&  39.9(6.3)&  5.9(3.1)& 0.127(5)&  0.126(4)& 52007.3924   &  6.1015690 & 0.440(0.017) & 197.6 (11.8) & $>$ 1000   \\
 TYC 8217-789-1  & 87.38(0.44)&  6000 &  5874(139)&  58.8(2.3)&  41.2(2.1)&  0.0     & 0.105(4)&  0.092(3)& 53452.8919   &  6.7867621 & 0.178(0.020) & 315.0 (7.2)  & $>$ 1000   \\
 TYC 9432-1633-1 & 88.27(0.50)&  5500 &  5770(227)&  35.2(4.1)&  64.8(5.4)&  0.0     & 0.054(5)&  0.065(6)& 53512.0943   & 10.0220862 & 0.319(0.095) & 327.2 (17.0) & $>$ 1000   \\
 SS TrA          & 89.43(0.72)&  7500 &  7424(156)&  63.4(1.9)&  36.6(1.6)&  0.0     & 0.077(2)&  0.060(2)& 53509.9548   &  8.6012129 & 0.519(0.029) & 216.6 (8.3)  & $>$ 1000   \\
 KO Nor          & 89.65(0.18)&  7300 &  6964(87) &  57.1(7.0)&  42.9(6.7)&  0.0     & 0.053(6)&  0.050(5)& 53485.9247   & 33.5618767 & 0.359(0.030) & 262.6 (1.9)  & $>$ 1000   \\
 V883 Sco        & 82.11(1.10)& 17000 & 15537(92) &  59.7(0.8)&  26.4(3.1)& 13.9(4.2)& 0.268(4)&  0.196(9)& 53120.8310   &  4.3411841 & 0.082(0.007) & 334.3 (1.0)  & 90.1 (2.3) \\
 V1301 Sco       & 83.65(0.15)&  6100 &  6340(36) &  29.5(0.6)&  70.5(0.6)&  0.0     & 0.123(2)&  0.177(3)& 56500.2186   &  1.9540710 & 0.183(0.003) & 118.8 (2.3)  & 435.3 (210)\\
 HD 158801       & 88.01(0.23)&  8000 &  8281(130)&  37.4(0.9)&  62.6(0.8)&  0.0     & 0.053(3)&  0.065(3)& 53505.6755   & 12.7764409 & 0.197(0.004) & 190.9 (21.5) & $>$ 1000   \\
 TYC 6258-1011-1 & 87.78(0.61)&  9000 &  9533(108)&  44.5(1.2)&  55.5(0.9)&  0.0     & 0.153(3)&  0.159(5)& 53644.0728   &  3.3156733 & 0.292(0.005) &  27.8 (18.3) & 273.2 (180)\\
 HD 163735       & 87.24(0.17)&  6700 &  6566(44) &  71.1(0.7)&  28.9(0.5)&  0.0     & 0.129(2)&  0.086(2)& 52002.7823   &  4.0501532 & 0.163(0.017) & 131.8 (22.7) & $>$ 1000   \\
 HD 313631       & 87.14(0.59)& 24000 & 20592(780)&  68.8(5.3)&  31.2(4.8)&  0.0     & 0.100(2)&  0.079(4)& 53545.4322   &  9.9214123 & 0.330(0.005) &  73.2 (1.0)  & $>$ 1000   \\
 HD 164610       & 86.98(0.36)&  8000 &  7496(100)&  57.1(2.6)&  40.3(1.4)&  2.6(1.8)& 0.064(3)&  0.062(6)& 52006.7671   &  7.8961010 & 0.029(0.010) & 204.1 (12.8) & 206.2 (112)\\
 V1344 Her       & 88.63(0.41)&  6000 &  5998(107)&  58.5(7.4)&  41.5(6.2)&  0.0     & 0.121(4)&  0.102(3)& 55501.8489   &  7.1461026 & 0.070(0.006) & 340.0 (19.4) & $>$ 1000   \\
 HD 170749       & 83.26(0.30)&  9500 & 11599(226)&  25.9(1.5)&  50.0(1.0)& 24.1(1.2)& 0.103(3)&  0.116(4)& 53501.2998   &  3.6898553 & 0.467(0.028) &  64.3 (10.6) & 791.4 (520)\\
 TYC 8378-252-1  & 87.65(0.92)&  6000 &  6077(144)&  53.4(1.6)&  46.6(1.2)&  0.0     & 0.153(6)&  0.134(5)& 51981.7745   &  2.8776861 & 0.169(0.018) & 157.4 (11.9) & 557.7 (341)\\
 TYC 6303-308-1  & 87.67(0.17)&  7000 &  6815(92) &  60.4(1.3)&  39.6(2.0)&  0.0     & 0.076(2)&  0.066(2)& 53499.4058   & 10.7037742 & 0.348(0.006) & 235.2 (7.8)  & $>$ 1000   \\
 PS Vul          & 79.51(1.48)& 14000 & 13005(493)&  22.2(3.1)&   7.8(2.7)& 70.0(10.3)&0.231(5)&  0.145(4)& 48504.8122   &  3.8173592 & 0.169(0.020) & 193.5 (18.0) & 280.0 (127)\\
 V839 Cep        & 86.59(0.39)& 12000 & 11445(157)&  52.1(2.0)&  47.7(1.6)&  0.2(0.2)& 0.067(3)&  0.066(3)& 51448.7078   &  9.9633587 & 0.037(0.019) &  60.5 (22.5) & $>$ 1000   \\
 TYC 5195-11-1   & 89.25(0.51)&  6500 &  6346(87) &  50.8(1.7)&  47.5(0.9)&  1.7(0.8)& 0.085(3)&  0.085(3)& 53508.2286   &  8.3587149 & 0.272(0.020) & 244.0 (8.7)  & $>$ 1000   \\
 TYC 2712-1201-1 & 80.71(0.16)& 10000 & 10296(110)&  29.7(0.4)&  43.7(0.6)& 26.6(3.0)& 0.167(2)&  0.195(2)& 51430.5593   &  1.8779673 & 0.196(0.009) &  57.6 (1.0)  & 86.8 (6.3) \\
UCAC4 585-123180 & 86.11(0.14)&  5900 &  5565(132)&  62.9(1.8)&  37.0(2.0)&  0.0     & 0.067(2)&  0.060(2)& 57045.7780   &  7.9515013 & 0.248(0.012) & 183.8 (12.3) & $>$ 1000   \\
 V922 Cep        & 86.86(0.09 &  6800 &  6838(77) &  48.9(2.0)&  49.6(1.8)&  1.5(1.2)& 0.098(2)&  0.098(2)& 51606.7539   &  3.5749727 & 0.135(0.005) &   2.3 (11.6) & 609.2 (450)\\
 V389 And        & 88.76(0.57)& 10000 &  9431(220)&  54.2(0.9)&  43.9(2.0)&  1.9(1.5)& 0.033(2)&  0.032(2)& 56203.7313   & 25.7783229 & 0.020(0.002) & 331.8 (1.8)  & $>$ 1000   \\
 \hline
 \end{tabular}}
\end{table*}

\section{Period eccentricity relation} \label{P-e}

\begin{figure}
  \centering
  \includegraphics[width=0.50\textwidth]{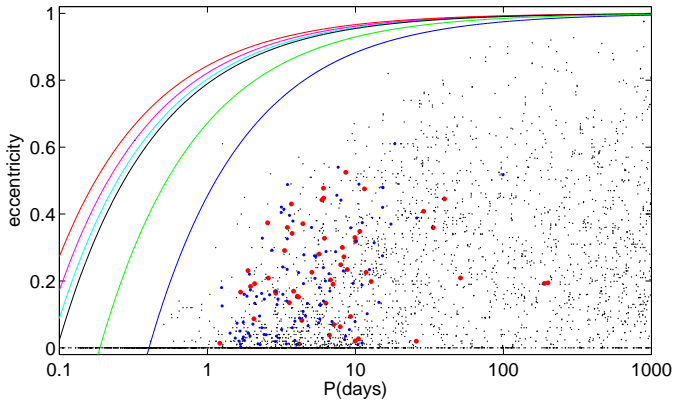}
  \caption{Distribution of known eccentric systems in the period-eccentricity diagram. New data are
  plotted in red, systems from the catalogue by \cite{2007MNRAS.378..179B} in blue, and spectroscopic
  binaries from the SB9 catalogue \citep{2004A&A...424..727P} as small black points. See the text for
  detailed description of the solid lines.}
  \label{FigP-e}
\end{figure}

The diagram which should be discussed in more detail is the period-eccentricity distribution of our
systems. We plotted our results together with those already published before by
\cite{2007MNRAS.378..179B} and also with the spectroscopic binaries from the SB9 catalogue by
\citep{2004A&A...424..727P} in the Fig. \ref{FigP-e}. As can be seen, orbital circularization plays
a role in binaries with shorter periods (\citealt{2010ApJS..190....1R},
\citealt{2013ARA&A..51..269D}), while practically all systems that have periods below one day are already
circular. Besides the plotted observed data we have also plotted several solid lines representing
the limits of very close periastron approaches of both components (i.e. 1.5$\times$ R$_\star$) when
these likely collide with each other. These periastron distances were computed for different
spectral types (B to M) according to their typical radii and masses (see
\citealt{1990hsaa.book.....Z}) with the assumption that both components are similar to each other
(same masses and radii).
As we can see, the sample of eclipsing apsidal motion systems has increased significantly with our
new data set and should be very helpful in future when analysing the P-e diagram in detail.

However, we still have to be very careful when interpreting these results with our new data. For
some systems a significant change of eccentricity could appear for the case when the star still has
poor data coverage of both light curve and the $O-C$ diagram. The P-e diagram still lacks of such
systems which have significant eccentricity close to its upper limit for a particular period. The
system with the highest eccentricity from the catalogue by \cite{2007MNRAS.378..179B} is LV~Her
(P=18.44d, e=0.61), while some more eccentric eclipsing systems definitely exist in the catalogue
of OGLE (Zasche, in prep.), and Kepler (Pr\v{s}a, priv.comm.).

 \section{Discussion and conclusions}

We have derived the preliminary apsidal motion and light curve parameters for 54 Algol-type
binaries. This is the first time any such analysis of such a large sample of eccentric eclipsing
binaries has been studied in our Galaxy using different sources of photometry. Bringing together
data from various databases and surveys has facilitated estimation of the long-term evolution of
the orbit and the apsidal precession of our sample. \cite{2016MNRAS.460..650H} is the only similar
study of such a large sample of stars. The authors  present 90 eclipsing binaries with apsidal
motion. However, their study only used the OGLE III database \citep{2011AcA....61..103G} covering
eight seasons. The advantage of our approach for the galactic targets is the fact that we also used
archival photometry and the already published data reach back even to the 1930s in one case
(SS~TrA), hence the detected slow apsidal motion should be more conclusive.

We should also mention the difference between our sample and the one already published
earlier by \cite{2007MNRAS.378..179B} -- both in magnitude ranges as well as spectral types. The
assumed spectral type and the distribution of both periods (orbital and apsidal motion) can also be
studied as noted for example, by \cite{2016MNRAS.460..650H}. They presented a diagram showing that there is
a possible relation between both periods and the masses of the components. However, they concluded
that a much larger sample is needed for the final verdict, especially the longer-periodic binaries
(P $>$ 6~d) that also have longer apsidal motions. In our contribution to the topic more than 50\% of
our studied systems have periods longer than six~days.

The sample of 54 systems presented in this study provides a good starting point for future
dedicated observations and analyses of several more interesting systems, for example, these showing
some suspicious additional variability of their orbital periods, these with rapid apsidal motion,
or those containing close components. As good spectroscopic data (i.e. knowing the individual
masses) is also available for them, one should be able to derive the internal structure constants
or compute the relativistic contribution to the total apsidal motion rate.

\begin{acknowledgements}

GJW gratefully thanks the Leverhulme Trust for financial support through a Fellowship. We would
like to thank the Pierre Auger Collaboration for the use of its facilities. The operation of the
robotic telescope FRAM was supported by the EU grant GLORIA (No. 283783 in FP7-Capacities programme)
and by the grants of the Ministry of Education of the Czech Republic (MSMT-CR LM2015038 and
LTT18004). The data calibration and analysis related to FRAM telescope is supported by the Ministry
of Education of the Czech Republic MSMT-CR CZ.02.1.01/0.0/0.0/16\_013/0001402. The STEREO HI
instrument featured in this study was developed by a collaboration that included the Rutherford
Appleton Laboratory and the University of Birmingham, both in the United Kingdom, the Centre
Spatial de Liege (CSL), Belgium, and the US Naval Research Laboratory (NRL),Washington DC, USA. The
STEREO/SECCHI project is an international consortium of the Naval Research Laboratory (USA),
Lockheed Martin Solar and Astrophysics Laboratory (USA), NASA Goddard Space Flight Center (USA),
Rutherford Appleton Laboratory (UK), University of Birmingham (UK), Max-Planck-Institut fur
Sonnensystemforschung (Germany), Centre Spatial de Liege (Belgium), Institut d'Optique Theorique et
Appliquee (France) and Institut d'Astrophysique Spatiale (France). This paper makes use of data
from the DR1 of the WASP data \citep{2010A&A...520L..10B} as provided by the WASP consortium, and
the computing and storage facilities at the CERIT Scientific Cloud, reg. no. CZ.1.05/3.2.00/08.0144
which is operated by Masaryk University, Czech Republic. Work is based on the data from the OMC
Archive at CAB (INTA-CSIC), pre-processed by ISDC. We thank the ASAS, NSVS, SuperWASP, OMC, and
ASAS-SN teams for making all of the observations easily public available. This work was supported
by the Czech Science Foundation grant no. GA15-02112S. Mr.Burkhardt from University of Heidelberg
is also greatly acknowledged for sending us the old scanned publications. This research has made
use of the SIMBAD database, operated at CDS, Strasbourg, France, and of NASA's Astrophysics Data
System Bibliographic Services.
\end{acknowledgements}

\hrule \vspace{2mm}

\textbf{Note added in proof:} A new study on eccentric eclipsing binaries by
\cite{2018ApJS..235...41K} was published while this manuscript was in the final stages of
publication. For five of the systems in common between \cite{2018ApJS..235...41K} and our paper,
the results are slightly different. This is likely to be due to different methodologies and
possibly also the different weightings used for the times of minima.

 \begin{appendix}
 \section{Appendix}\label{apendix}

\begin{figure*}[h]
  \centering
  \includegraphics[width=0.92\textwidth]{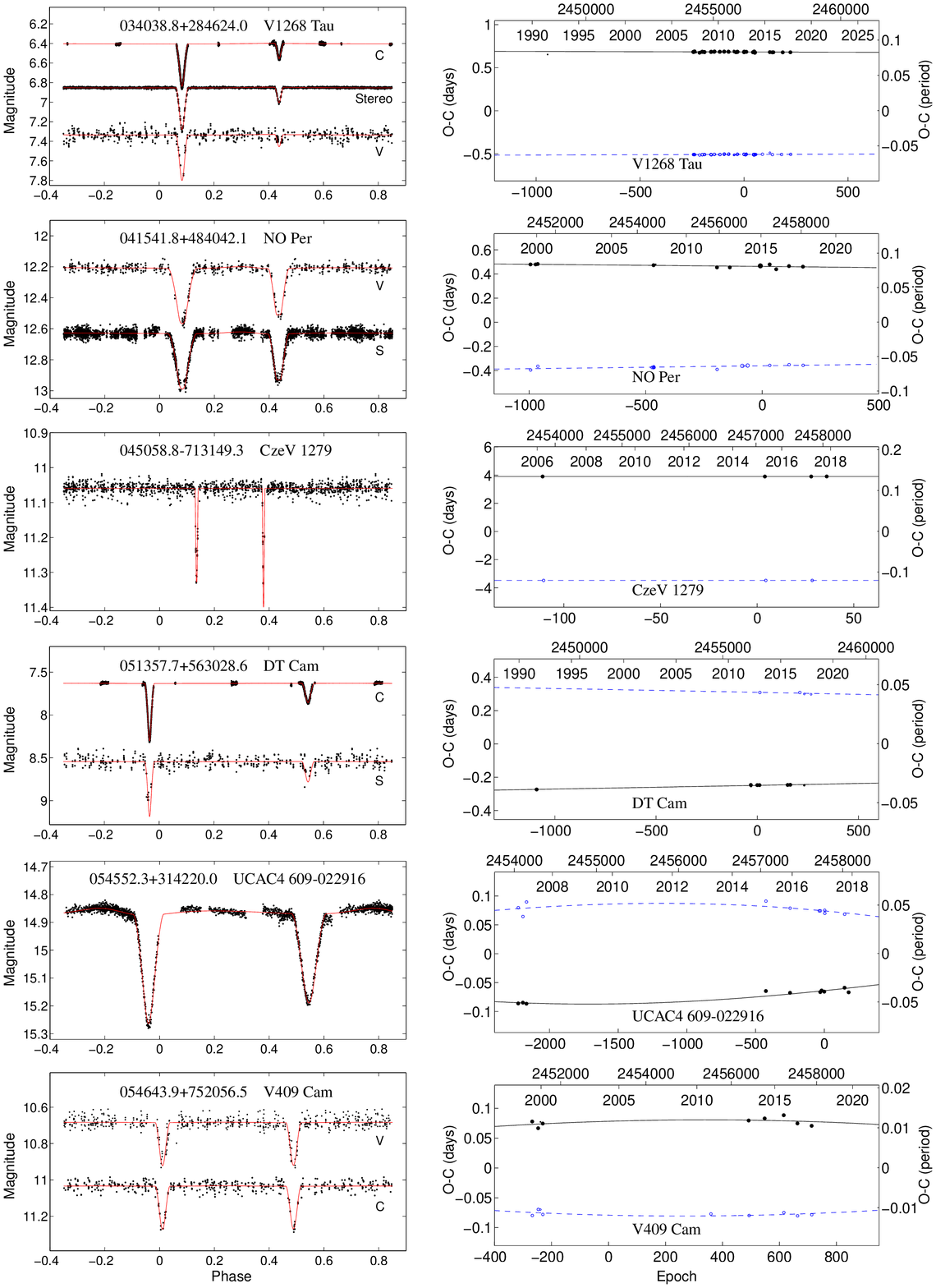}
  \caption{Plot of the light curves and $O-C$ diagrams of the analysed systems, cont.}
  \label{FigLCOC2}
\end{figure*}

\begin{figure*}[h]
  \centering
  \includegraphics[width=0.92\textwidth]{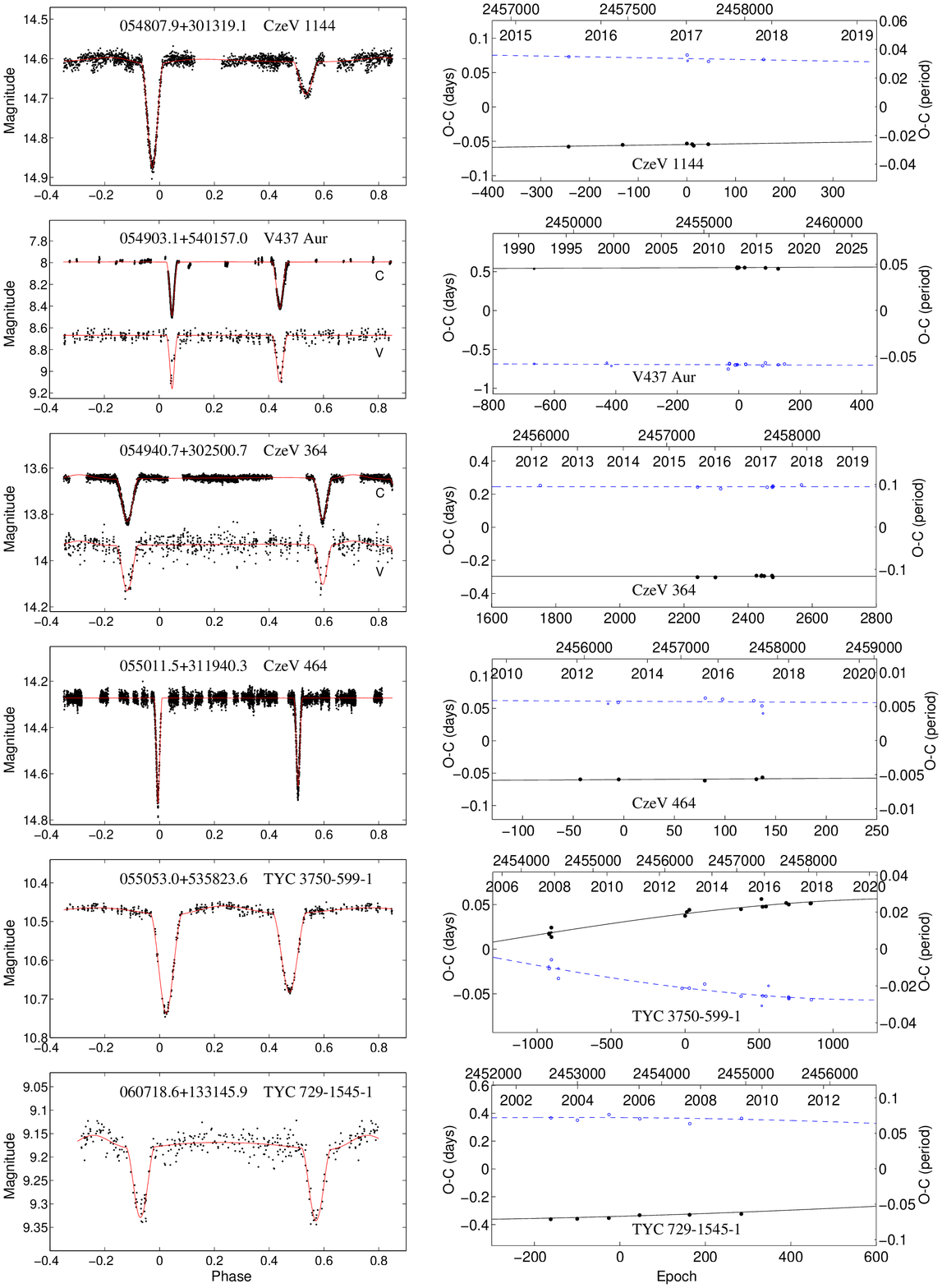}
  \caption{Plot of the light curves and $O-C$ diagrams of the analysed systems, cont.}
  \label{FigLCOC3}
\end{figure*}

\begin{figure*}[h]
  \centering
  \includegraphics[width=0.92\textwidth]{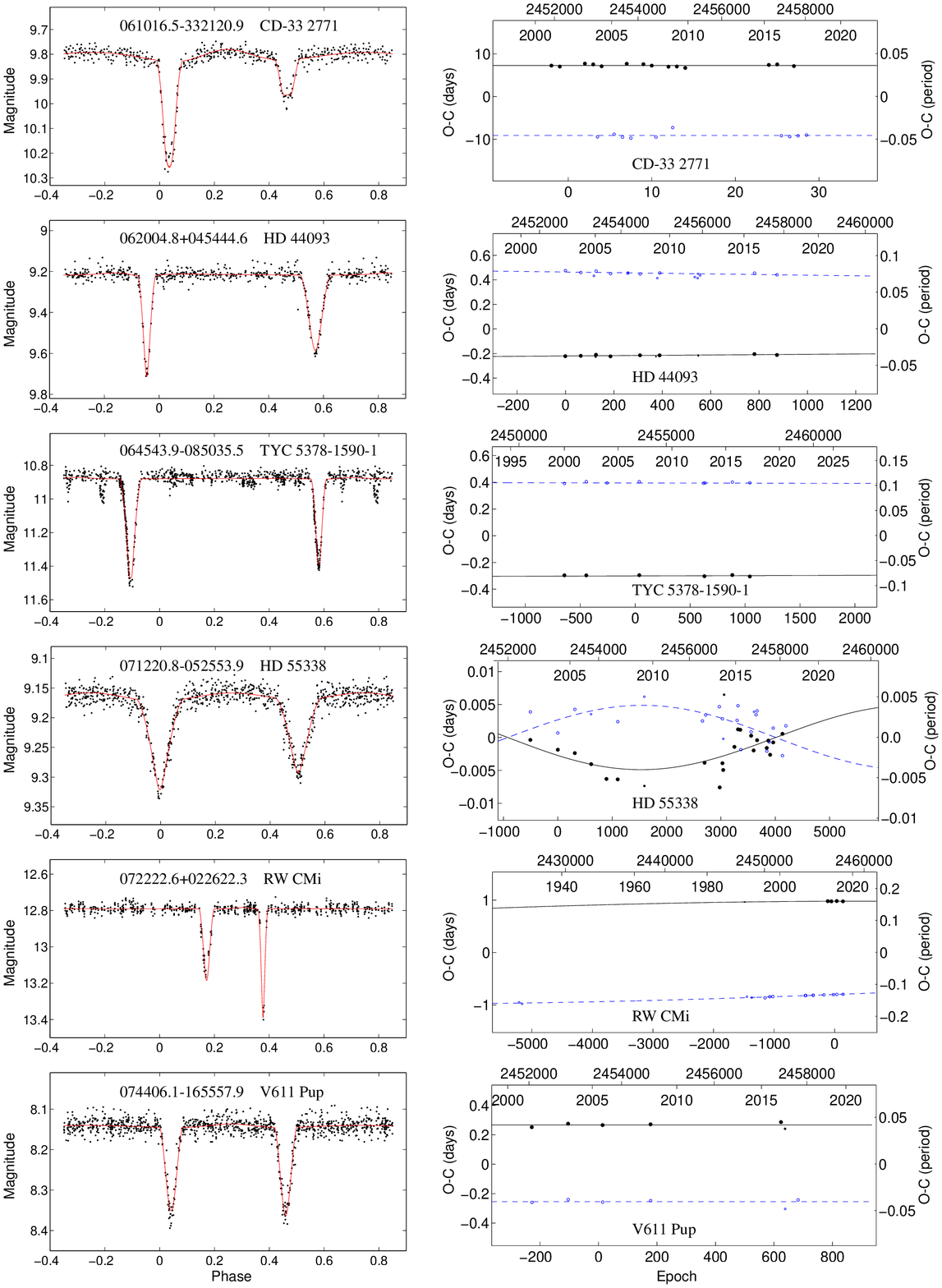}
  \caption{Plot of the light curves and $O-C$ diagrams of the analysed systems, cont.}
  \label{FigLCOC4}
\end{figure*}

\begin{figure*}[h]
  \centering
  \includegraphics[width=0.92\textwidth]{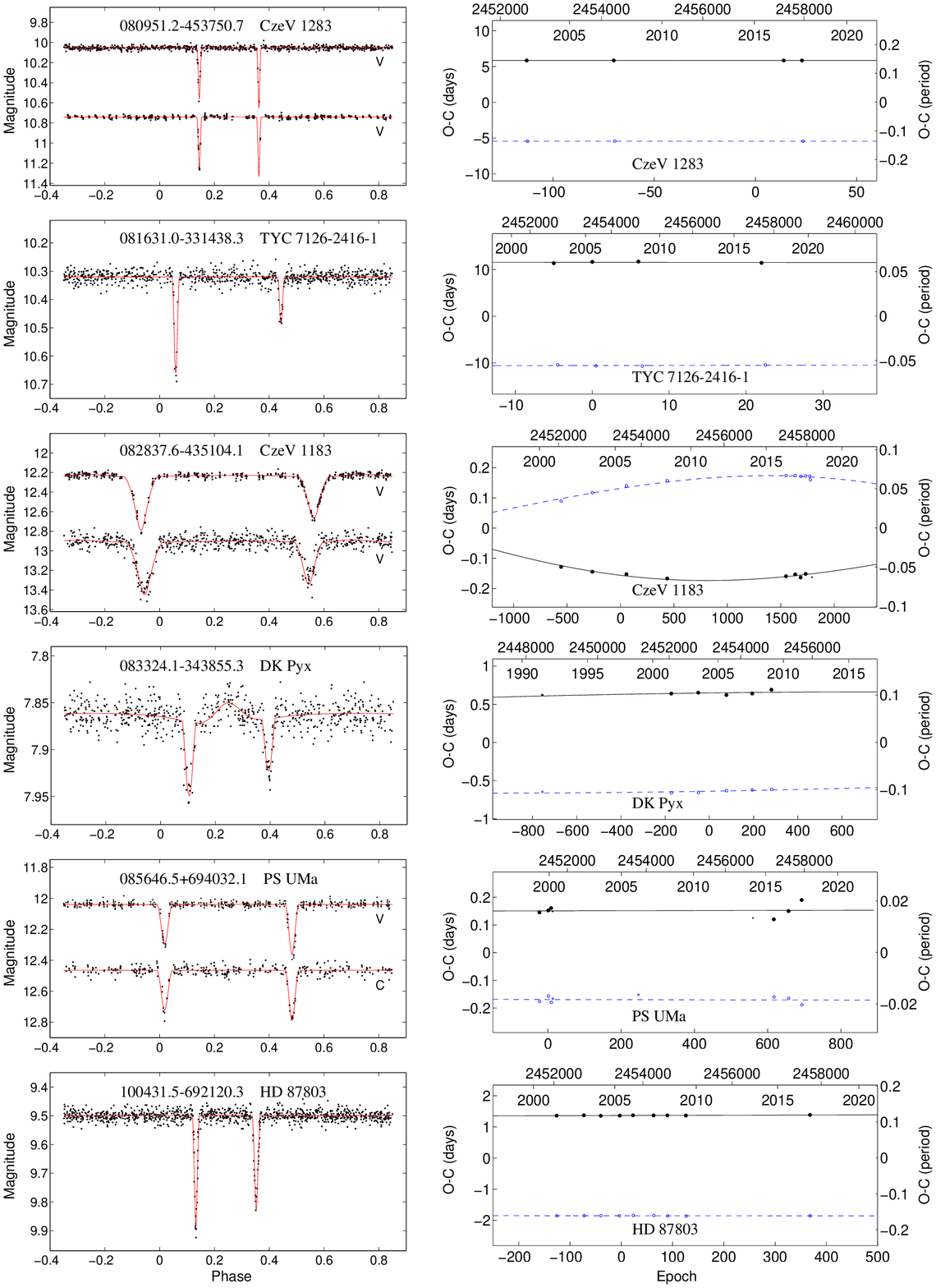}
  \caption{Plot of the light curves and $O-C$ diagrams of the analysed systems, cont.}
  \label{FigLCOC5}
\end{figure*}

\begin{figure*}[h]
  \centering
  \includegraphics[width=0.92\textwidth]{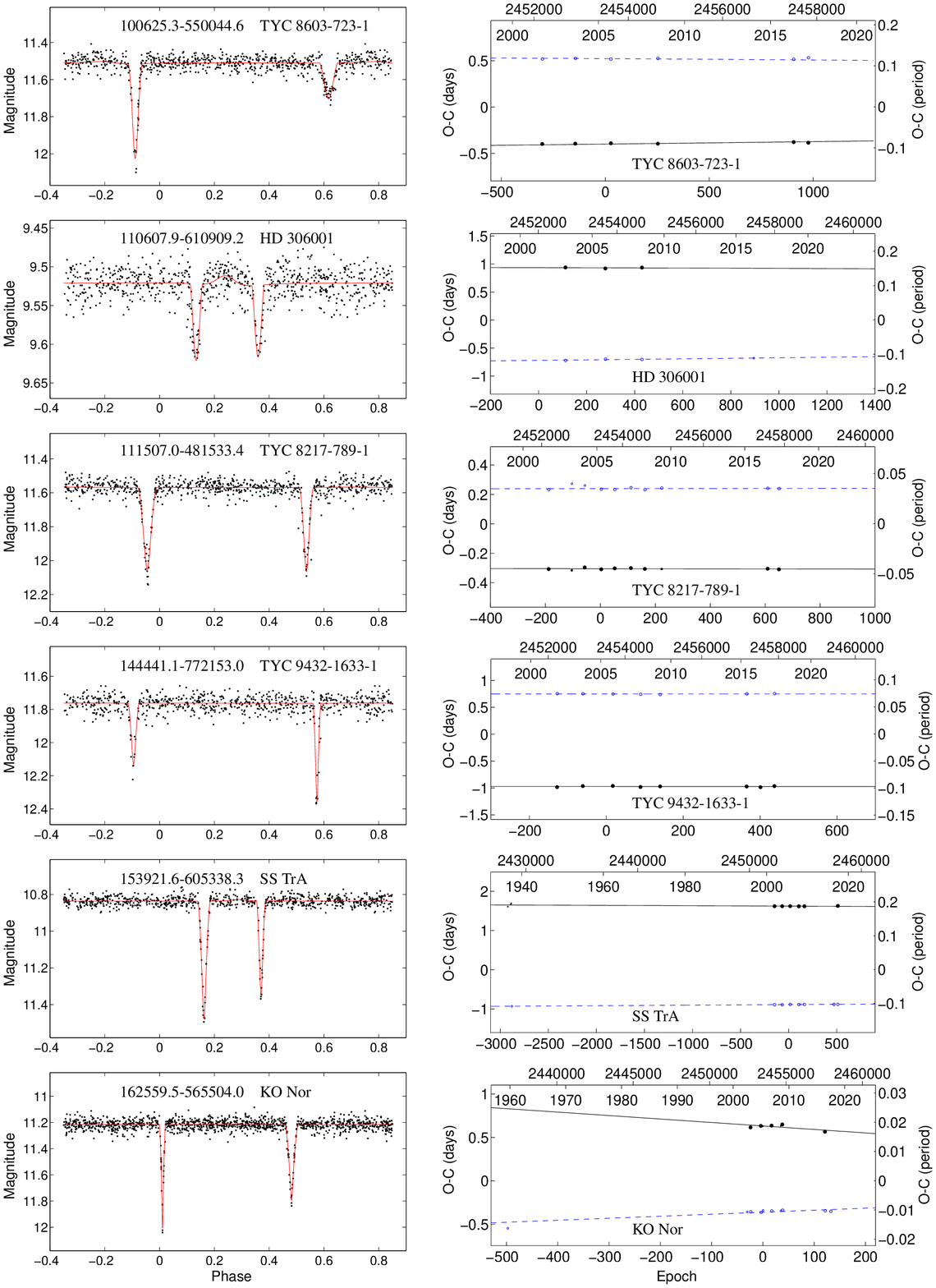}
  \caption{Plot of the light curves and $O-C$ diagrams of the analysed systems, cont.}
  \label{FigLCOC6}
\end{figure*}

\begin{figure*}[h]
  \centering
  \includegraphics[width=0.92\textwidth]{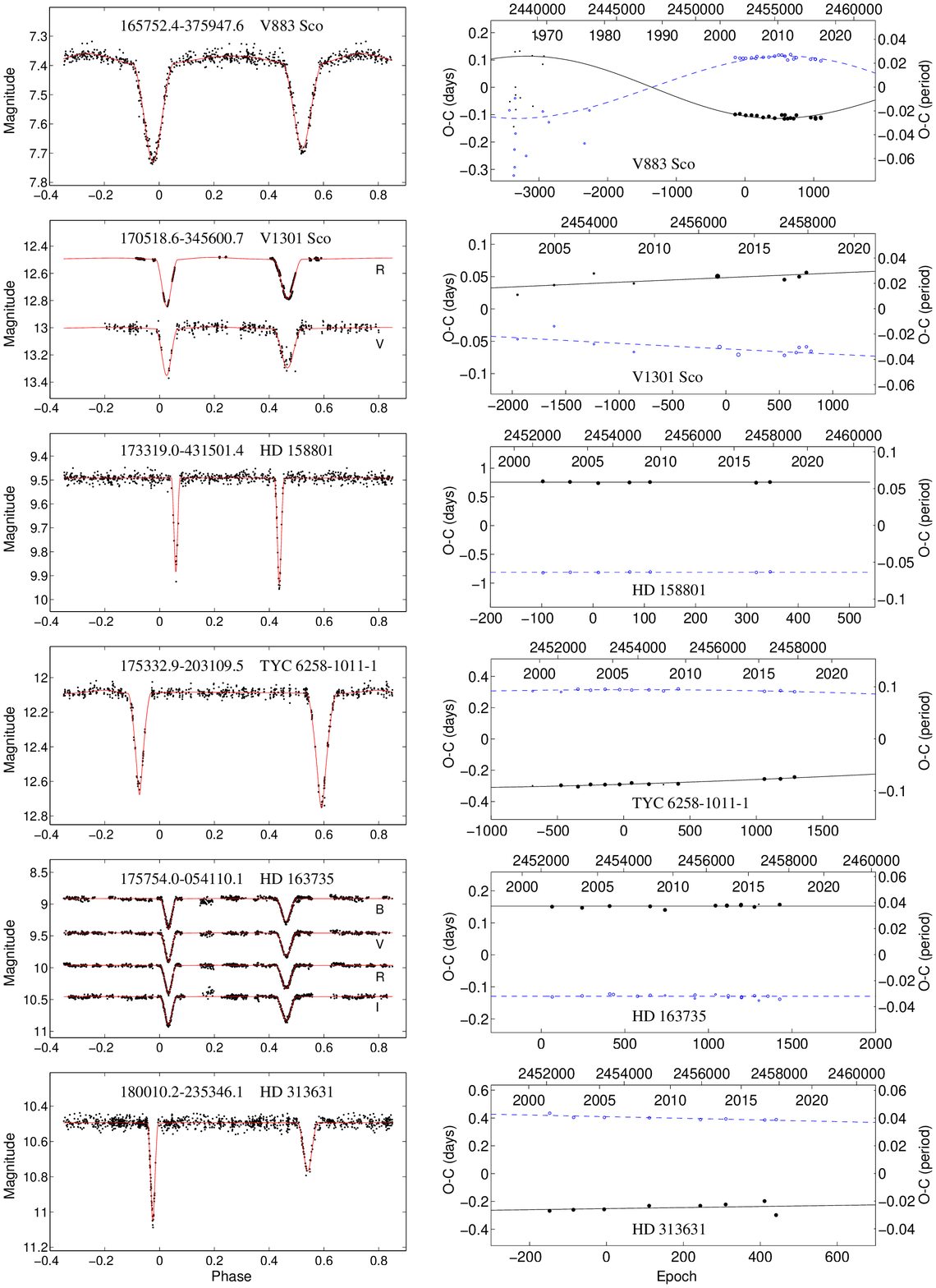}
  \caption{Light curves and $O-C$ diagrams of the analysed systems, cont.}
  \label{FigLCOC7}
\end{figure*}

\begin{figure*}[h]
  \centering
  \includegraphics[width=0.92\textwidth]{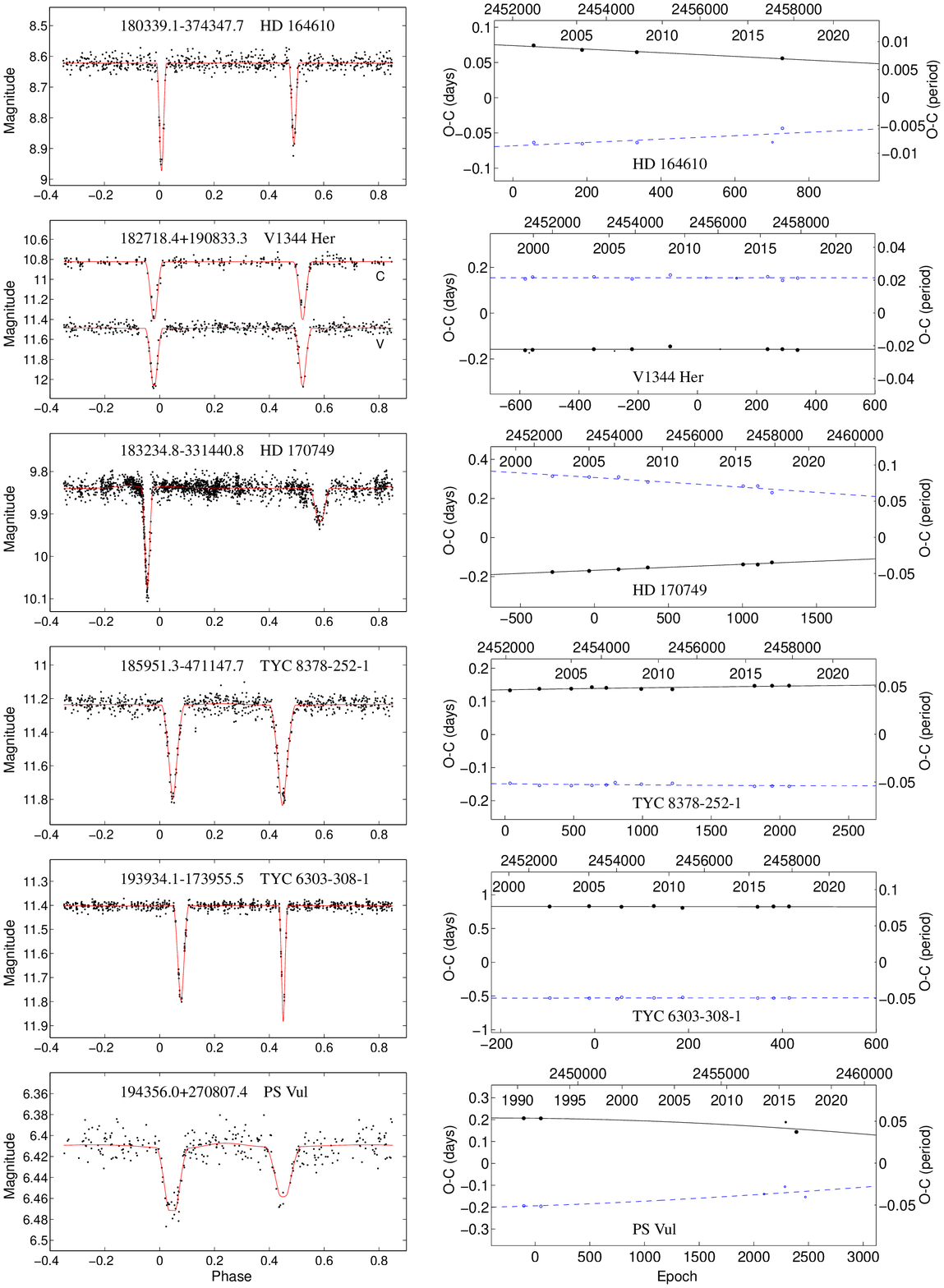}
  \caption{Light curves and $O-C$ diagrams of the analysed systems, cont.}
  \label{FigLCOC8}
\end{figure*}

\begin{figure*}[h]
  \centering
  \includegraphics[width=0.92\textwidth]{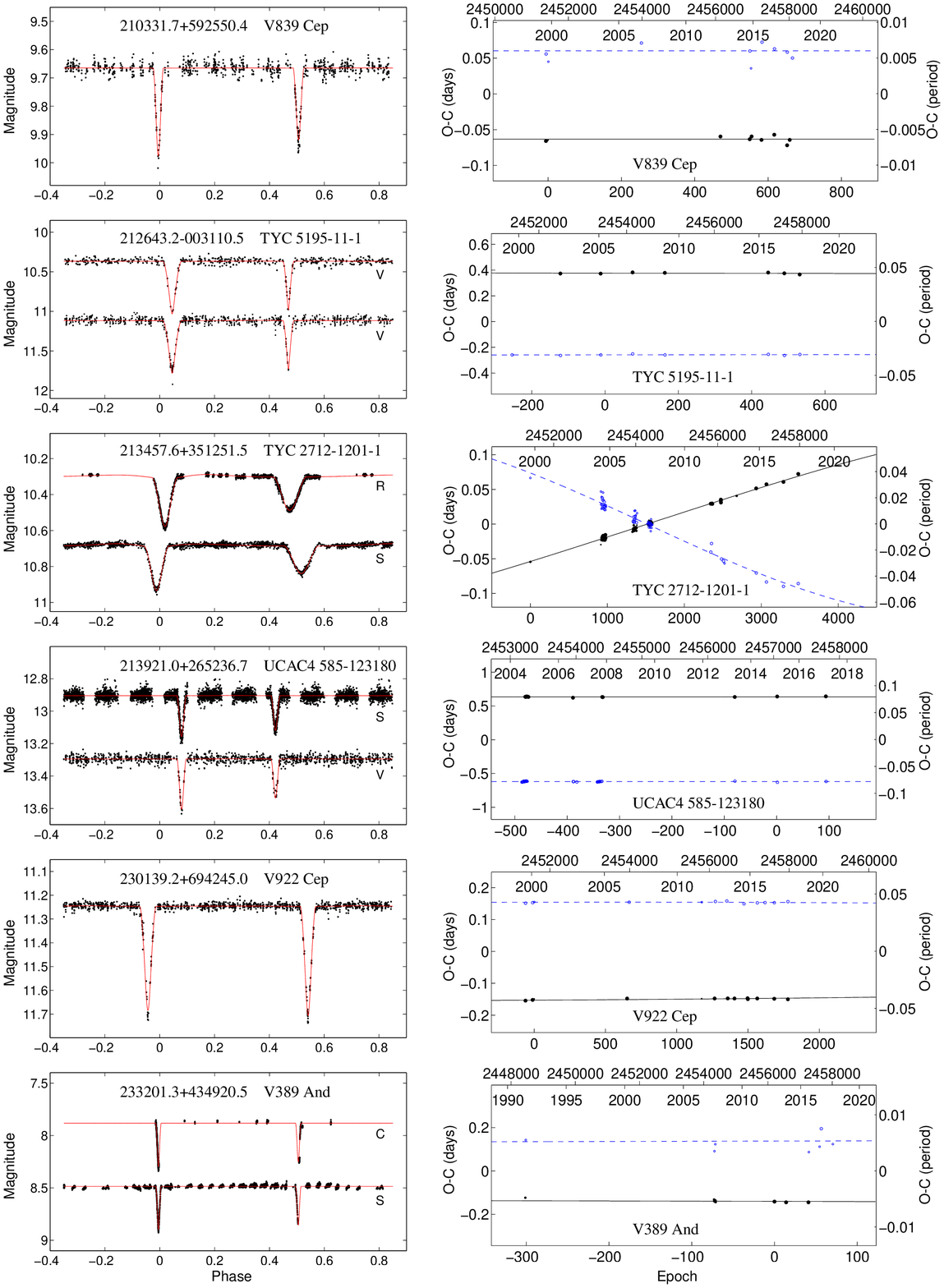}
  \caption{Light curves and $O-C$ diagrams of the analysed systems, cont.}
  \label{FigLCOC9}
\end{figure*}

\begin{table}
 \caption{Apsidal rotators with significant relativistic contribution.}
 \label{TabRelativistic}
 \centering
\begin{tabular}{l l}
 \hline\hline
 Star  & $\dot\omega_{rel}/\dot\omega_{total}$\\
 \hline
 V1268 Tau& 59.9 $\pm$ 35.0 \% \\
 V437 Aur & 37.5 $\pm$ 14.7 \% \\
 HD 44093 & 52.8 $\pm$ 18.4 \% \\
 V611 Pup & 73.2 $\pm$ 40.8 \% \\
 HD 163735& 22.3 $\pm$ 15.7 \% \\
 HD 313631& 57.1 $\pm$ 23.5 \% \\
 V1344 Her& 55.0 $\pm$ 42.8 \% \\
 \hline \hline
\end{tabular}
\end{table}

\newpage

 \section{Tables with minima}\label{minima}

\begin{table*}
 \caption{Heliocentric minima used for the analysis.}
 \label{Minima}
 \tiny
 \centering
 \begin{tabular}{l l l c c l}
 \hline\hline
 Star  & HJD - 2400000 & Error  & Type & Filter & Reference\\
 \hline
 V1137 Cas & 51473.568   &        & P & V & IBVS 5570 \\
 V1137 Cas & 53613.4050  & 0.0002 & S & R & IBVS 5741 \\
 V1137 Cas & 53613.40344 & 0.0002 & S & R & OEJV 74 \\
 V1137 Cas & 53615.45611 & 0.0003 & P & R & New - This study \\
 V1137 Cas & 53746.48711 & 0.0009 & S & R & New - This study \\
 V1137 Cas & 54763.33788 & 0.0001 & P & R & OEJV 107 \\
 V1137 Cas & 55470.36914 & 0.0001 & P & R & New - This study \\
\hline
 \hline \hline
\end{tabular}
\end{table*}

\begin{table*}
 \caption{Heliocentric minima used for the analysis, cont.}
 \tiny
 \centering
 \begin{tabular}{l l l c c l}
 \hline\hline
 Star  & HJD - 2400000 & Error  & Type & Filter & Reference\\
 \hline
 V1018 Cas & 51514.9605  & 0.00727& P & V & NSVS - This study \\
 V1018 Cas & 51516.6615  & 0.00669& S & V & NSVS - This study \\
 V1018 Cas & 51601.6250  &        & P &   & IBVS 5586 \\
 V1018 Cas & 54833.7032  & 0.0018 & P & V & IBVS 5894 \\
 V1018 Cas & 55543.6722  & 0.0011 & P & V & IBVS 5960 \\
 V1018 Cas & 55832.6117  & 0.00033& P & R & IBVS 6007 \\
 V1018 Cas & 55883.811   & 0.008  & S & V & IBVS 6011 \\
\hline \hline
\end{tabular}
\end{table*}

\begin{table*}
 \caption{Heliocentric minima used for the analysis, cont.}
 \tiny
 \centering
 \begin{tabular}{l l l c c l}
 \hline\hline
 Star  & HJD - 2400000 & Error  & Type & Filter & Reference\\
 \hline
 V1268 Tau & 55597.36479 & 0.00154& S & St& STEREO - This study \\
 V1268 Tau & 55602.63508 & 0.00223& P & St& STEREO - This study \\
 V1268 Tau & 55605.52991 & 0.00189& S & St& STEREO - This study \\
 V1268 Tau & 55806.67715 & 0.01830& P & St& STEREO - This study \\
 V1268 Tau & 55809.54847 & 0.11881& S & St& STEREO - This study \\
 V1268 Tau & 55814.82353 & 0.04043& P & St& STEREO - This study \\
 V1268 Tau & 55817.71300 & 0.02531& S & St& STEREO - This study \\
\hline \hline
\end{tabular}
\end{table*}

\begin{table*}
 \caption{Heliocentric minima used for the analysis, cont.}
 \tiny
 \centering
 \begin{tabular}{l l l c c l}
 \hline\hline
 Star  & HJD - 2400000 & Error  & Type & Filter & Reference\\
 \hline
 UCAC4 609-022916 & 57065.19325 & 0.00090 & S & V & ASAS-SN - This study \\
 UCAC4 609-022916 & 57358.75521 & 0.00057 & P & V & ASAS-SN - This study \\
 UCAC4 609-022916 & 57359.73871 & 0.00076 & S & V & ASAS-SN - This study \\
 UCAC4 609-022916 & 57726.95356 & 0.00117 & P & V & ASAS-SN - This study \\
 UCAC4 609-022916 & 57727.93118 & 0.00033 & S & V & ASAS-SN - This study \\
 UCAC4 609-022916 & 58023.19244 & 0.00062 & P & V & ASAS-SN - This study \\
 UCAC4 609-022916 & 58024.15657 & 0.00126 & S & V & ASAS-SN - This study \\
\hline \hline
\end{tabular}
\end{table*}

\begin{table*}
 \caption{Heliocentric minima used for the analysis, cont.}
 \tiny
 \centering
 \begin{tabular}{l l l c c l}
 \hline\hline
 Star  & HJD - 2400000 & Error  & Type & Filter & Reference\\
 \hline
 CzeV 464         & 57781.46458 & 0.00855 & P & C & New - This study \\
 CzeV 464         & 57839.20344 & 0.00289 & S & C & New - This study \\
 CzeV 464         & 57247.11431 & 0.00341 & P & V & ASAS-SN - This study \\
 CzeV 464         & 57252.48045 & 0.00496 & S & V & ASAS-SN - This study \\
 CzeV 464         & 57844.33190 & 0.00359 & P & V & ASAS-SN - This study \\
 CzeV 464         & 57849.66937 & 0.01451 & S & V & ASAS-SN - This study \\ \hline
 TYC 3750-599-1   & 56238.38854 & 0.00224 & S & V & New - This study \\
\hline \hline
\end{tabular}
\end{table*}

\begin{table*}
 \caption{Heliocentric minima used for the analysis, cont.}
 \tiny
 \centering
 \begin{tabular}{l l l c c l}
 \hline\hline
 Star  & HJD - 2400000 & Error  & Type & Filter & Reference\\
 \hline
 HD 44093         & 54470.39657 & 0.00261 & S & V & ASAS - This study  \\
 HD 44093         & 54948.02316 & 0.00234 & P & V & ASAS - This study  \\
 HD 44093         & 54951.66513 & 0.00497 & S & V & ASAS - This study  \\
 HD 44093         & 53335.55875 & 0.23281 & S & C & Pi of the sky - This study \\
 HD 44093         & 53361.64026 & 0.02023 & P & C & Pi of the sky - This study \\
 HD 44093         & 53379.45626 & 0.05675 & P & C & Pi of the sky - This study \\
 HD 44093         & 54173.33398 & 0.00816 & S & C & Pi of the sky - This study \\
\hline \hline
\end{tabular}
\end{table*}

\begin{table*}
 \caption{Heliocentric minima used for the analysis, cont.}
 \tiny
 \centering
 \begin{tabular}{l l l c c l}
 \hline\hline
 Star  & HJD - 2400000 & Error  & Type & Filter & Reference\\
 \hline
 RW CMi           & 50533.415   & 0.007   & S &   & BBSAG 115 \\ 
 RW CMi           & 50837.609   & 0.003   & S &   & BBSAG 119 \\ 
 RW CMi           & 54092.461   & 0.005   & S & C & OEJV 73   \\ 
 RW CMi           & 54153.2996  & 0.0009  & S & I & IBVS 5802 \\ 
 RW CMi           & 54889.4443  & 0.0001  & S &   & JAAVSO 38 \\ 
 RW CMi           & 56354.3940  & 0.0039  & P & I & IBVS 6084 \\ 
 RW CMi           & 56713.3366  & 0.0037  & P & I & IBVS 6118 \\
\hline \hline
\end{tabular}
\end{table*}

\begin{table*}
 \caption{Heliocentric minima used for the analysis, cont.}
 \tiny
 \centering
 \begin{tabular}{l l l c c l}
 \hline\hline
 Star  & HJD - 2400000 & Error  & Type & Filter & Reference\\
 \hline
 PS UMa           & 57937.67613 & 0.00349 & S & V & ASAS-SN - This study \\
 PS UMa           & 53802.54968 & 0.00218 & S & R & New - This study \\
 PS UMa           & 56700.22225 & 0.00799 & P & R & New - This study \\ \hline
 HD 87803         & 52062.22504 & 0.00475 & P & V & ASAS - This study \\
 HD 87803         & 52672.29739 & 0.00462 & P & V & ASAS - This study \\
 HD 87803         & 53052.13108 & 0.00650 & P & V & ASAS - This study \\
 HD 87803         & 53478.02607 & 0.00195 & P & V & ASAS - This study \\
\hline \hline
\end{tabular}
\end{table*}

\begin{table*}
 \caption{Heliocentric minima used for the analysis, cont.}
 \tiny
 \centering
 \begin{tabular}{l l l c c l}
 \hline\hline
 Star  & HJD - 2400000 & Error  & Type & Filter & Reference\\
 \hline
 SS TrA           & 52248.85705 & 0.00049 & S & V & ASAS - This study \\
 SS TrA           & 52928.35222 & 0.00570 & S & V & ASAS - This study \\
 SS TrA           & 53650.86277 & 0.00110 & S & V & ASAS - This study \\
 SS TrA           & 54416.36841 & 0.00107 & S & V & ASAS - This study \\
 SS TrA           & 54915.24066 & 0.00545 & S & V & ASAS - This study \\
 SS TrA           & 57564.41757 & 0.00107 & S & V & ASAS-SN - This study \\
 SS TrA           & 57906.67723 & 0.00177 & P & V & ASAS-SN - This study \\
\hline \hline
\end{tabular}
\end{table*}

\begin{table*}
 \caption{Heliocentric minima used for the analysis, cont.}
 \tiny
 \centering
 \begin{tabular}{l l l c c l}
 \hline\hline
 Star  & HJD - 2400000 & Error  & Type & Filter & Reference\\
 \hline
 V883 Sco         & 39671.958   &         & P &   & Strohmeier \& Knigge (1973) \\ 
 V883 Sco         & 39710.844   &         & P &   & Strohmeier \& Knigge (1973) \\ 
 V883 Sco         & 39954.203   &         & P &   & Strohmeier \& Knigge (1973) \\ 
 V883 Sco         & 40027.999   &         & P &   & Strohmeier \& Knigge (1973) \\ 
 V883 Sco         & 40336.128   &         & P &   & Strohmeier \& Knigge (1973) \\ 
 V883 Sco         & 40338.126   &         & S &   & Strohmeier \& Knigge (1973) \\ 
 V883 Sco         & 40409.958   &         & P &   & Strohmeier \& Knigge (1973) \\
\hline \hline
\end{tabular}
\end{table*}

\begin{table*}
 \caption{Heliocentric minima used for the analysis, cont.}
 \tiny
 \centering
 \begin{tabular}{l l l c c l}
 \hline\hline
 Star  & HJD - 2400000 & Error  & Type & Filter & Reference\\
 \hline
 HD 163735        & 56838.79611 & 0.00036 & P &BVRI& FRAM - This study \\
 HD 163735        & 56842.84780 & 0.00045 & P &BVRI& FRAM - This study \\
 HD 163735        & 56852.68301 & 0.00113 & S &BVRI& FRAM - This study \\
 HD 163735        & 57166.85316 & 0.00028 & P &BVRI& FRAM - This study \\
 HD 163735        & 57168.60084 & 0.00039 & S &BVRI& FRAM - This study \\
 HD 163735        & 57492.61155 & 0.00038 & S &BVRI& FRAM - This study \\
 HD 163735        & 57272.16528 & 0.00468 & P & V & ASAS-SN - This study \\
\hline \hline
\end{tabular}
\end{table*}

\begin{table*}
 \caption{Heliocentric minima used for the analysis, cont.}
 \tiny
 \centering
 \begin{tabular}{l l l c c l}
 \hline\hline
 Star  & HJD - 2400000 & Error  & Type & Filter & Reference\\
 \hline
 TYC 8378-252-1   & 55484.05502 & 0.00151 & P & C & CRTS - This study \\
 TYC 8378-252-1   & 55485.20987 & 0.00105 & S & C & CRTS - This study \\
 TYC 8378-252-1   & 57204.92178 & 0.00032 & P & V & ASAS-SN - This study \\
 TYC 8378-252-1   & 57206.05708 & 0.00100 & S & V & ASAS-SN - This study \\
 TYC 8378-252-1   & 57573.26581 & 0.00041 & P & V & ASAS-SN - This study \\
 TYC 8378-252-1   & 57574.40176 & 0.00107 & S & V & ASAS-SN - This study \\
 TYC 8378-252-1   & 57930.09913 & 0.00071 & P & V & ASAS-SN - This study \\
\hline \hline
\end{tabular}
\end{table*}

\begin{table*}
 \caption{Heliocentric minima used for the analysis, cont.}
 \tiny
 \centering
 \begin{tabular}{l l l c c l}
 \hline\hline
 Star  & HJD - 2400000 & Error  & Type & Filter & Reference\\
 \hline
 TYC 2712-1201-1  & 53166.76175 & 0.00270 & S & S & WASP - This study  \\
 TYC 2712-1201-1  & 53167.65770 & 0.00018 & P & S & WASP - This study  \\
 TYC 2712-1201-1  & 53169.53653 & 0.00289 & P & S & WASP - This study  \\
 TYC 2712-1201-1  & 53170.53471 & 0.00102 & S & S & WASP - This study  \\
 TYC 2712-1201-1  & 53181.79158 & 0.00265 & S & S & WASP - This study  \\
 TYC 2712-1201-1  & 53182.68400 & 0.00155 & P & S & WASP - This study  \\
 TYC 2712-1201-1  & 53183.66692 & 0.00021 & S & S & WASP - This study  \\
\hline \hline
\end{tabular}
\end{table*}

\begin{table*}
 \caption{Heliocentric minima used for the analysis, cont.}
 \tiny
 \centering
 \begin{tabular}{l l l c c l}
 \hline\hline
 Star  & HJD - 2400000 & Error  & Type & Filter & Reference\\
 \hline
 TYC 2712-1201-1  & 56080.43853 & 0.00031 & P & R & New - This study  \\
 TYC 2712-1201-1  & 56094.44192 & 0.00086 & S & R & New - This study  \\
 TYC 2712-1201-1  & 56141.38808 & 0.00103 & S & R & New - This study  \\
 TYC 2712-1201-1  & 56167.67619 & 0.00588 & S & R & New - This study  \\
 TYC 2712-1201-1  & 56463.55363 & 0.00317 & P & R & New - This study  \\
 TYC 2712-1201-1  & 56940.56850 & 0.00027 & P & R & New - This study  \\
 TYC 2712-1201-1  & 56941.38541 & 0.00016 & S & R & New - This study  \\
\hline
 \hline \hline
\end{tabular}
\end{table*}

 \end{appendix}

\end{document}